\documentclass[twocolumn]{aastex701}

\usepackage{graphicx}
\usepackage{amsmath}
\usepackage{hyperref}
\usepackage{makecell}
\usepackage{array}
\usepackage{tabularx}
\usepackage{multirow}
\usepackage{balance}

\newcommand{\Mdot}{{\rm M}_{\odot}}
\newcommand{\kms}{\mathrm{km~s^{-1}}}
\newcommand{\hh}{h^{-1}\ }

\begin{document}

\title{Identification of Compact Groups of Galaxies in IllustrisTNG300}

\author[0009-0000-1346-4777]{Seungwu Yoo}
\email{yoosw2001@snu.ac.kr}
\affiliation{Astronomy Program, Department of Physics and Astronomy, Seoul National University, Gwanak-gu, Seoul 08826, Korea}

\author[0000-0002-9254-144X]{Jubee Sohn}
\affiliation{Department of Physics and Astronomy, Seoul National University, 1 Gwanak-ro, Gwanak-gu, Seoul 08826, Republic of Korea}
\affiliation{SNU Astronomy Research Center, Seoul National University, Seoul 08826, Republic of Korea}
\email{jubee.sohn@snu.ac.kr}

\begin{abstract}
We identify compact groups of galaxies (CGs) in the IllustrisTNG-300 simulation using a Friends-of-Friends (FoF) algorithm. Our approach is designed to be comparable to systematic CG searches based on spectroscopic surveys, while avoiding the conventional Hickson selection criteria \citep{Hickson1982}, which can bias samples toward relatively low-density environments. We construct two CG catalogs: one based on a three-dimensional distance linking length of 73 kpc (i.e., $50~h^{-1}$ kpc), and another based on projected and radial linking lengths of 73 kpc and $1000~\kms$. We refer to these as the position–position–position (PPP) and position–position–velocity (PPV) CG catalogs, respectively. The PPV catalog provides a direct analog to observed CG samples. At $z = 0$ in TNG300, we identify 383 PPP CGs and 1666 PPV CGs.  A large fraction ($\sim 80\%$) of PPV CGs are not physically compact systems but are contaminated by line-of-sight interlopers. We demonstrate that the scaling relation between total group stellar mass and velocity dispersion is an effective diagnostic for identifying false positives with line-of-sight interlopers. We further examine the large-scale environments of CGs and show that they reside in a wide range of densities, including the central regions of galaxy clusters. These CG catalogs provide a robust foundation for studying the formation and evolution of CGs in cosmological simulations. 
\end{abstract}

\section{Introduction} \label{sec:intro}

Compact groups of galaxies (hereafter CGs) are associations of typically three to ten galaxies confined within an extremely compact physical scale of only a few tens of kpc. Their galaxy number densities are remarkably high, often exceeding even the number density of the cores of rich galaxy clusters (the median value of galaxy number density of $\log(\rho/(\rm Mpc^3))\sim3.6$, \citealp{Sohn2016}). The velocity dispersion between members of CGs are also generally very low ($\sim 200~\kms$). This combination of dense concentration and low relative velocities implies that CGs provide ideal environments for frequent gravitational interactions among member galaxies. Thus, CGs have been regarded as ideal laboratories for studying the role of galaxy–galaxy encounters in driving galaxy evolution (e.g., \citealp{VerdesMontenegro1998, VerdesMontenegro2001, Rasmussen2008, Walker2013, Bitsakis2016, Lee2017}). If such interactions are indeed common, the member galaxies in a CG are expected to merge into a single massive remnant on relatively short timescales, comparable to the crossing time of groups ($\sim 1\text{-}2$ Gyr; \citealp{Carnevali1981, Barnes1989}).

The high observed abundance of CGs, despite their predicted short dynamical lifetimes, remains a long standing puzzle (e.g., \citealp{Carnevali1981, Barnes1989, Governato1991, Diaferio1994, Athanassoula1997}). The simplest explanation for the abundance of CGs is that a substantial fraction of CGs are not physically bound systems but chance alignments of galaxies along the line of sight (e.g., \citealp{Mamon1986}). The large spectroscopic survey of CG members enables a quantitative evaluation of contamination by chance alignments (e.g., \citealp{Hickson1992, McConnachie2009, Mendel2011, Sohn2015}), suggesting that a large fraction of CGs indeed consist of member galaxies at similar redshifts. Furthermore, the detection of X-ray emitting gas in CGs \citep{Ebeling1994, Ponman1996} indicates that CGs are genuine concentration of galaxies with heated intragroup gas. Later, studies based on numerical simulations suggest that CG configurations can survive longer than the merging time scale when the mass ratio between the member galaxies are large \citep{Governato1991} or CG members are within a compact massive common halo \citep{Athanassoula1997}. 

Another proposed explanation for the observed abundance of CGs is that member galaxies merge into a single massive galaxy in a short time scale, while new galaxies from the surrounding environment are continuously accreted, thereby regenerating CG configurations (e.g., \citealp{Diaferio1994}). If CGs are continuously formed, the surrounding density environment becomes a key factor in determining their lifetimes. In dense environments, CG configurations may be sustained over extended periods, whereas isolated CGs are expected to disperse on much shorter timescales. Motivated by this connection between environment and CG lifetime, many observational studies have investigated the abundance of CGs as a function of redshift \citep{McConnachie2009, Sohn2015, Sohn2016, Zheng2020, DiazGimenez2018, Zandivarez2022, Zandivarez2024}.

The very first step for understanding the abundance and the lifetime of CGs is constructing the catalog of CGs. The Hickson Compact Group (HCG) Catalog \citep{Hickson1982} is the pioneering attempt for a systematic search for CGs. \citet{Hickson1982} introduced a well-established definition of CGs (see Section \ref{sec:CGdefinition}) and identified 100 HCGs based on the Palomar Observatory Sky Survey (POSS). Since \citet{Hickson1982}, many studies construct the CG catalogs based on the Hickson criteria applied to various observational datasets \citep{Prandoni1994, Iovino2002, Lee2004, McConnachie2009}. \citet{Hickson1992} refined the HCG catalog based on spectroscopy; the chance alignments and the line-of-sight contamination are removed using the radial velocity measurements. Similarly, the large spectroscopic surveys including SDSS were used for refining various CG catalogs (e.g., \citealp{Iovino2003, Lee2004, deCarvalho12005, McConnachie2009, Mendel2011, Pompei2012, DiazGimenez2012, DiazGimenez2018, Sohn2015, Zheng2020, Zandivarez2022, Zandivarez2024}).

Despite the widespread use in identifying CGs, the Hickson criteria introduce systematic biases that can significantly affect subsequent analyses of CG physical properties. The isolation criterion, which indicates that there should be no galaxy with comparable luminosity to the CG member may lie within an annulus spanning one to three times the group radius, was originally introduced by \citet{Hickson1982} to avoid selecting systems embedded in cluster cores. However, this density-dependent selection leads to biased interpretations of CG lifetimes across different environments. In particular, if CGs are continually replenished by surrounding galaxies to maintain their compact configurations, excluding systems in dense regions preferentially introduces a systematic bias in CG properties (e.g., \citealp{Barton1996, Sohn2016}). 

To mitigate this selection effect, \citet{Barton1996} identified CGs based on a Friends-of-Friends (FoF) algorithm \citep{Huchra1982}. CGs identified with the FoF algorithm show physical properties that differ from those selected with the Hickson criteria. Similarly, \citet{Sohn2016} also identified CGs using the FoF algorithm applied to a much larger spectroscopic sample of Sloan Digital Sky Survey (SDSS). \citet{Sohn2016} demonstrated that CGs selected without the isolation criterion show distinct physical characteristics, including larger sizes and higher velocity dispersions, compared to their counterparts identified with the traditional Hickson criteria.

Although observational studies offer insights into the nature and evolution of CGs, fully tracing their evolutionary pathways requires theoretical guidance from numerical simulations. For example, cosmological simulations including IllustrisTNG \citep{Nelson2018, Pillepich2018, Naiman2018, Marinacci2018, Springel2018} we use here offer statistically large samples of CGs across various redshift range enabling detailed investigation of their physical properties and evolutionary trend. In this work, we first build the CG catalog from IllustrisTNG for a comprehensive study for CGs in cosmological simulation while maintaining consistency with observational CG identification. We particularly use the FoF approach to construct the CG catalog in IllustrisTNG. This method complements many previous simulation-based studies of CGs (e.g., \citealp{DiazGimenez2020, Hartsuiker2020, Taverna2024, FloresFreitas2024, Celiz2025}), which predominantly rely on the Hickson criteria. Our goal is to assess the performance of FoF based CG identification in terms of completeness and purity, and establish a reliable framework for connecting simulated CGs with their observational counterparts. 

We organize this paper as follows. We first describe IllustrisTNG simulation we used in Section \ref{sec:TNG}. We demonstrate the CG identification in Section \ref{sec:CGIdentification}. We examine the physical properties of CGs in Section \ref{sec:catalogs} and compare the TNG CGs with SDSS CGs in Section \ref{sec:CG_Obs_comparison}. We then estimate the environmental densities of CGs and investigate their properties across different environments in Section \ref{sec:embedded_cg}. We conclude in Section \ref{sec:Conclusion}. Throughout the paper, we adopt cosmological parameters $\Omega_{\Lambda} = 0.6911$, $\Omega_{m} = 0.3089$, $H_0=67.74~\mathrm{km~s^{-1}~Mpc^{-1}}$. 

\section{ILLUSTRIS-TNG} \label{sec:TNG}

We study CGs in IllustrisTNG, a set of cosmological magnetohydrodynamical simulations for studying the formation and evolution of galaxies and large-scale structures \citep{Nelson2018, Pillepich2018, Naiman2018, Marinacci2018, Springel2018}. In particular, we use IllustrisTNG-300 (hereafter TNG300) that covers the largest volume of $300\ \mathrm{Mpc}^{3}$ among a suite of the IllustrisTNG simulations. The large volume of TNG300 enables the identification of a large number of CGs. TNG300 consists of three simulations with various mass resolutions. We use IllustrisTNG-300-1 with the best mass resolution of the dark matter particle mass $m_{DM} = 5.9\times10^7\ \Mdot$ and the baryon particle mass $m_{baryon} = 1.1\times10^7\ \Mdot$. Hereafter, we also refer to IllustrisTNG-300-1 as TNG300. 

We use the subhalo catalog in a $z=0$ snapshot of TNG300, including subhalos identified with the \texttt{SUBFIND} algorithm \citep{Springel2001}. The TNG subhalo catalog includes the three-dimensional positions, velocities, the total mass (i.e., the sum of dark matter, gas, stellar particle, and blackholes within each subhalo), and a half-mass size of all subhalos. 

We select subhalos with a stellar mass larger than $10^{9}\ \Mdot$. The mass limit we apply roughly corresponds to the stellar mass limit of various dense spectroscopic surveys \citep{Sohn2017}. We also choose the subhalos with cosmic origin (i.e., \texttt{SubhaloFlag} $= 1$). The subhalos with \texttt{SubhaloFlag} $= 0$ are either satellite halos formed within one virial radius of their parent halos or the dark matter mass is less than 80\% of the total mass of subhalos \citep{Nelson2019}. We basically exclude these subhalos for identifying CGs and discuss the impact of the omission of these subhalos in Section \ref{sec:subhaloflag}.

\section{Compact Group Identification} \label{sec:CGIdentification}

\subsection{Compact Group Definition} \label{sec:CGdefinition}

CGs are the densest galaxy systems, consisting of three to ten galaxies confined within a region spanning only a few tens of kiloparsecs. The most widely adopted definition of CGs is introduced by \citet{Hickson1982} (hereafter the Hickson criteria):
\begin{enumerate}
\item Population: $N_{G} \geq 4$,
\item Isolation: $\theta_{N} > 3\theta_{G}$,
\item Compactness: $\bar{\mu}_{G} < 26$ mag arcsec$^{-2}$.
\end{enumerate}
Here, $N_{G}$ indicates the number of member galaxies that are no more than three magnitudes fainter than the brightest group galaxy. $\theta_{G}$ is the radius of the smallest circle enclosing all group members, and $\theta_{N}$ is the radius of the largest concentric circle free of any non-member galaxies. $\bar{\mu}_{G}$ is the mean surface brightness of the member galaxies within $\theta_{G}$. These criteria have been widely applied to identify CGs in both observational datasets (e.g., \citealp{Hickson1982, Prandoni1994, Iovino2002, Iovino2003, Lee2004, deCarvalho12005, McConnachie2009, Mendel2011, Pompei2012, DiazGimenez2012, DiazGimenez2018, Sohn2015, Zheng2020, Zandivarez2022, Zandivarez2024}) and cosmological simulations (e.g., \citealp{DiazGimenez2020, Hartsuiker2020, Taverna2024, Celiz2025}).

The isolation criterion was introduced to exclude systems that are embedded within larger-scale structures (i.e., galaxy clusters). However, this criterion introduces a selection bias by excluding CGs located in dense environments \citep{Barton1996, Sohn2016}. As a result, CG catalogs constructed strictly based on the Hickson criteria may be incomplete and introduce selection biases in terms of CG abundance and evolutionary pathways of CGs. Indeed, \citet{Sohn2016} demonstrated that the physical properties of CG, including size and the number density, differ significantly from those of CG candidates identified based only on the Hickson selection. We do not use the Hickson CG selection criteria, and we follow the methodology of \citet{Sohn2016} to construct a more complete CG catalog in TNG300.

\subsection{The Friends-of-Friends Algorithm} \label{sec:FoF}

We apply the Friends-of-Friends algorithm (hereafter, the FoF algorithm, \citealp{Huchra1982}) to identify CGs in TNG300. The FoF algorithm identifies galaxy systems based on proximity. Starting from a single galaxy, the algorithm identifies galaxies closer than the linking length as friends. Then, the algorithm iterates the identification of all other friend galaxies of friend galaxies within the linking length and bundles them up into a single group. The FoF algorithm is widely applied to identify galaxy systems in both observations and simulations (e.g., \citealp{Robotham2011, Tempel2016, Pillepich2018, Sohn2021}). 

The choice of linking length determines the characters of identified systems. The FoF algorithm can be applied with a distance linking length or projected and radial linking lengths. The FoF algorithm with a distance linking length bundles systems with galaxies based on the three dimensional distance (particularly, when we know the three-dimensional (3D) distances as in numerical simulations). In contrast, the FoF algorithm with projected and radial linking lengths enables system identification in redshift space (e.g., based on spectroscopic surveys).

Identifying systems based on the FoF algorithm with the radial linking length is often suffering from false identification. A chance alignment with galaxies that are not located at the similar physical distance, but with similar radial velocities cannot be excluded based on the FoF algorithm applied to spectroscopy. Evaluation of contamination of chance alignment requires a full three-dimensional positional and velocity information of galaxies. Recent developments of cosmological numerical simulations enable this evaluation. Here, we thus test the identification of CGs based on the FoF algorithm with a distance linking length to identify 3D concentration of galaxies and with projected and radial linking lengths to identify concentration of galaxies in the redshift space, imitating the CG identification based on spectroscopic survey.

\subsection{Position–Position–Position (PPP) Compact Groups} \label{sec:3dcgcat}

We first identify CGs as systems of galaxies concentrated within a compact 3D volume. We apply the FoF algorithm with a single linking length of 73 kpc ($= 50~h^{-1}$ kpc) to the TNG300 subhalo sample, selecting objects with $M_{*} > 10^{9}~\Mdot$ and with \texttt{SubhaloFlag} $ =1$. The choice of linking length follows the projected linking length adopted in spectroscopic CG searches in SDSS \citep{Sohn2016}. In this procedure, subhalos separated by less than 73 kpc in 3D space are grouped into candidate systems. We then select FoF systems containing between three and ten member galaxies as CG candidates. We include triplets in our sample (and in the CG catalog), as their physical properties do not differ significantly from those of systems with four or more members \citep{Duplancic2013, Sohn2015}. We also account for the periodic boundary conditions of the simulation by recentering the galaxies used in the FoF identification at the center of the simulation box. There are 1428 CG candidates, hereafter, PPP (position-position-position) FoF groups, identified in TNG300.

We impose additional selection to identify CGs composed of galaxies with similar masses. Following the Hickson population criterion, member galaxies are no fainter than three magnitudes compared to the brightest group members. \citet{Hickson1982} applied this magnitude selection along with other selection criteria to exclude the cores of rich clusters. Similarly, many studies including \citet{Sohn2016} imposed a similar magnitude limit when they identified CGs based on the FoF algorithm to the spectroscopic survey. \citet{Sohn2016} reported that systems violate the magnitude difference limit contain one dominant bright galaxy and fainter satellites.

Following these observational studies, we identify CGs that comprise galaxies with similar mass. In simulations, the stellar mass is a more fundamental measure than the luminosity. Thus, we select the systems containing three to nine member galaxies with a stellar mass ratio between most and third massive member smaller than one-fifteenth $(M_{*,1}/M_{*,3}\ge15)$ corresponding to the three magnitudes with assumption that the luminosity of a galaxy is proportional to stellar mass. We finally identify 383 CGs out of 1428 PPP FoF groups; hereafter, we refer to these 383 systems as PPP CGs.

\begin{figure*}
\centering
\includegraphics[width=1\linewidth]{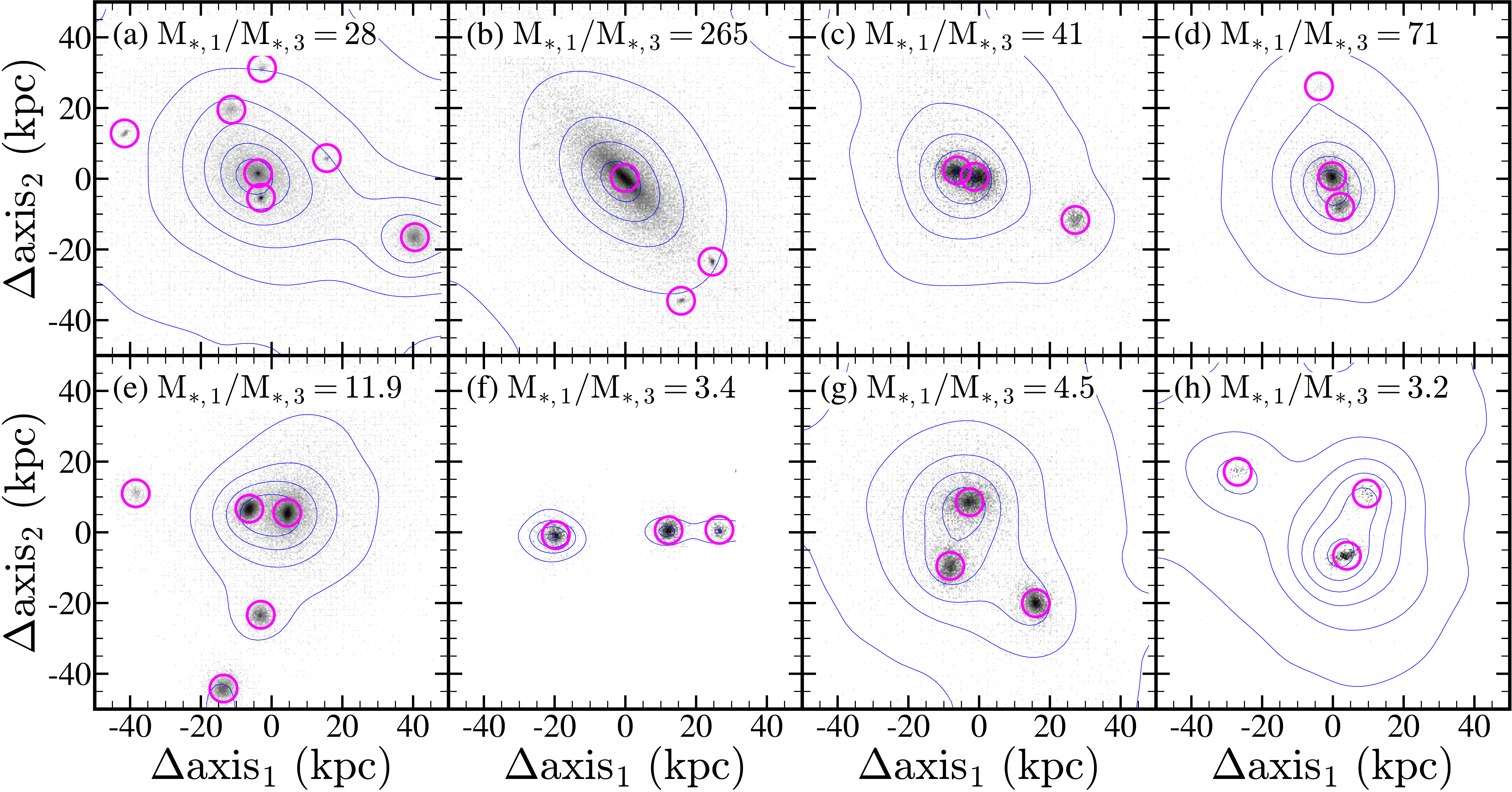}
\caption{(Upper panels) Spatial distributions of subhalos in CG candidates (i.e., FoF groups in our notation) consist of massive subhalo surrounded by satellite subhalos. The underlying gray density map shows the distribution of stellar particles and blue contours indicate the number density of dark matter particles. Magenta circles mark the location of member subhalos. We set different axes in each figure so that the separation of the member subhalos is distinctive. (Lower panels) Same as upper panels, but for PPP CGs consist of the members with similar masses. }
\label{fig:FOF_CG_examples}
\end{figure*}

Figure \ref{fig:FOF_CG_examples} displays the example CG candidates we identify from TNG300 and demonstrates the impact of the mass ratio selection. The top panels display the PPP FoF groups that do not satisfy the mass ratio selection. These systems consist of a dominant central galaxy surrounded by many satellite galaxies. The lower panels of Figure \ref{fig:FOF_CG_examples} show examples of PPP CGs; the member galaxies have similar masses comprising member galaxies with similar stellar masses. Hereafter, these PPP CGs are the main subject of our study. 

\subsection{Position-Position-Velocity (PPV) Compact Groups} \label{sec:prcgcat}

We next apply the FoF algorithm with projected and radial linking lengths to identify CGs mocking the observational approach. This approach is comparable to the CG identification based on spectroscopic surveys \citep{Barton1996, Sohn2016}. Here, we mock the observational CG identification by using the projected linking distance on the $x-y$ plane and the radial velocity difference in the $z-$direction. In other words, we simply assume that we observed the simulated universe on the $x-y$ plane. Following \citet{Sohn2016}, we use the FoF algorithm with a projected linking length of 73 kpc and a radial linking length of $1000~\kms$.

For a fair comparison with the observations, we also consider the cosmic expansion effect depending on the distance to galaxy subhalos in the $z-$direction. We first run the FoF algorithm to identify CG candidates. We then assume an imaginary observer is located at 100 Mpc along with the $z-$direction from each CG candidate. We consider the periodic boundary condition by shifting galaxies in the line-of-sight direction. We then compute the expansion velocity of galaxies based on: 
\begin{equation}
d(z_{exp})= \int_0^{z_{exp}}\frac{cdz'}{H(z')},
\end{equation}
where $H(z) = H_{0}\sqrt {\Omega_\Lambda + \Omega_m(1+z)^3}$. 
For the individual galaxies, we compute $z_{exp}$ using the $z-$direction distance, and convert them in to expansion velocities (i.e., $v_{exp} = cz_{exp}$). We then apply the expansion velocity to the simulated peculiar velocity to obtain the observational radial velocity of individual galaxies: $(1+z_{obs}) = (1 + v_{obs}/c) =(1+z_{exp})(1+z_{pec})$.

Finally, we apply the FoF algorithm using the same projected and radial linking lengths, but now based on the observationally motivated radial velocities. This procedure yields 1666 CGs in TNG300. Hereafter, we refer to these systems as PPV (position–position–velocity) CGs.

Among 1666 PPV CGs, only 338 systems are genuine PPP CGs (i.e., true positives). The number of genuine CGs among PPV CGs is smaller than the total number of PPP CGs. In other words, many PPV CGs are the systems of galaxies within similar radial velocities, but not concentrated in a compact region. There are 45 PPP CGs missing from the PPV CGs. Among these 45 missed PPP CGs, there are 30 missed PPP CGs because of additional members with similar radial velocities with similar radial velocities with PPP CG members (but not within 73 kpc, i.e., interlopers), and the inclusion of these interlopers violate the population criterion (either the mass ratio between members or the number of member galaxies in a CG). There are remaining 15 cases where PPP CG members have extremely large radial velocity differences, thus excluded from the PPV CG identification. Not all 338 true positives in PPV CGs are identical with the PPP CGs; 88 true positives include more than one members that are not physically concentrated, but within the line-of-sight velocities (i.e., $|\Delta V| < 1000~\kms$).

\section{The Catalogs of Compact Groups in TNG300} \label{sec:catalogs}

We construct PPP and PPV CG catalogs in TNG300, and Table~\ref{tab:CG_mem_num} summarizes the number of systems we identified. Using a sample that includes only cosmic-origin subhalos with $M_{*} > 10^{9}~\Mdot$, we initially identify 1428 PPP FoF groups and 4714 PPV FoF groups. Applying the stellar mass–ratio selection reduces these to 383 PPP CGs and 1666 PPV CGs. For comparison with observations, we also report the number of CGs identified from a mock volume-limited sample (see Section \ref{sec:CG_Obs_comparison}). In addition, we construct CG catalogs using a sample that includes both cosmic- and non-cosmic origin subhalos with $M_{*} > 10^{9}~\Mdot$ (see Section~\ref{sec:subhaloflag}). Finally, we also list the member galaxies in each CG sample.

\begin{deluxetable*}{lccccccc} 
\label{tab:CG_mem_num}
\tabletypesize{\footnotesize}
\tablecaption{Summary of TNG300 Compact Group Catalogs}
\tablehead{\colhead{Subhalo Sample} & \colhead{Group catalog} & \colhead{$\mathrm{N_{FoF}}$} & \colhead{$\mathrm{N_{CG}}$} & \colhead{$\mathrm{N_{V1 CG}}^{1}$}  & \colhead{$\mathrm{N_{Mem, FoF}}$} & 
\colhead{$\mathrm{N_{Mem, CG}}$}  & \colhead{$\mathrm{N_{Mem, V1 CG}}^{1}$}}
\startdata
$M_{*} > 10^{9}~\Mdot~\&$           & PPP CG$^{2}$ & 1428 &  383 &     229 &  4820 & 1265 &  720 \\
Only cosmic origin subhalos$^{4,5}$ & PPV CG$^{3}$ & 4714 & 1666 &     897 & 16001 & 6123 & 3113 \\
\hline
$M_{*} > 10^{9}~\Mdot~\&$           & PPP CG$^{2}$ & 2713 &  532 & \nodata & 10497 & 1872 & \nodata \\
Cosmic/non-cosmic origin subhalos$^{6}$ & PPV CG$^{3}$ & 5554 & 1853 & \nodata & 23604 & 7019 & \nodata \\
\enddata
\tablenotetext{1}{CGs identified from a volume-limited sample for comparison with observed CGs (see Section~\ref{sec:CG_Obs_comparison}).}
\tablenotetext{2}{Position–position–position (PPP) CGs identified using a three-dimensional linking length of 73 kpc.}
\tablenotetext{3}{Position–position–velocity (PPV) CGs identified using a projected linking length of 73 kpc and a radial linking length of $1000~\kms$.}
\tablenotetext{4}{Cosmic-origin subhalos have \texttt{SubhaloFlag} $=1$, while non-cosmic-origin subhalos have \texttt{SubhaloFlag} $=0$.}
\tablenotetext{5}{The primary sample used in this study.}
\tablenotetext{6}{Comparison sample constructed from a galaxy catalog that includes both cosmic and non-cosmic origin subhalos (see Section \ref{sec:subhaloflag}).}
\end{deluxetable*}

We also provide the member catalogs for the PPP CGs (Table \ref{tab:submem_3d}) and the PPV CGs (Table \ref{tab:submem_proj}). Based on the member subhalo properties, we further derive physical properties of two types of CGs, including the total mass and velocity dispersion (Tables \ref{tab:3dcg_pp} and \ref{tab:prcg_pp}). We explain how we compute these physical properties in Section \ref{sec:pp_measurement}. We also discuss an important scaling relation between the total stellar mass and velocity dispersion of CGs in Section \ref{sec:Msig}. 

\begin{deluxetable*}{ccccccccc}
\tablecaption{Members of PPP CGs in TNG300}
\label{tab:submem_3d}
\tablehead{
\colhead{\multirow{2}{*}{ID$_{\rm PPP}$}} &
\colhead{\multirow{2}{*}{Subhalo ID}} &
\colhead{$\log M_*$} &
\colhead{$x$} & \colhead{$y$} & \colhead{$z$} &
\colhead{$v_x$} & \colhead{$v_y$} & \colhead{$v_z$} \\
&
&
\colhead{($\mathrm{M_\odot}$)} &
\colhead{($\mathrm{kpc}$)} &
\colhead{($\mathrm{kpc}$)} &
\colhead{($\mathrm{kpc}$)} &
\colhead{($\kms$)} &
\colhead{($\kms$)} &
\colhead{($\kms$)}
}
\startdata
0 &  145 & 10.8 & 64613 & 72206 & 217808 &	  76 &  605	&  -671 \\
0 &  158 & 10.7 & 64562 & 72212 & 217810 &	-264 & -103 & -1630 \\
0 &  330 & 10.3 & 64562 & 72215 & 217826 &  -268 & 1559 & -1308 \\
1 &  613 & 9.9  & 66983 & 76432 & 216148 &  -115 & -960 &  1537 \\
1 & 1271 & 9.3  & 67032 & 76458 & 216179 &	-132 & -630 &  -475 \\
1 & 1496 & 9.0  & 67029 & 76379 & 216167 &  -698 & -700 &   -72 \\
2 & 894  & 9.6  & 64438 & 71917 & 217728 &   113 & -811 &  -800 \\
2 & 1142 & 9.3  & 64447 & 71889 & 217713 &	1689 & 2431 &   549 \\
2 & 1213 & 9.2  & 64389 & 71927 & 217708 &  1847 & 1926 &  1155 \\
\enddata
\tablecomments{This table is available in its entirety in machine-readable form in the online journal.}
\end{deluxetable*}

\begin{deluxetable*}{ccccccccc}
\label{tab:submem_proj}
\tablecaption{Members of PPV CGs in TNG300}
\tablehead{
\colhead{\multirow{2}{*}{ID$_{\rm PPV}$}} &
\colhead{\multirow{2}{*}{Subhalo ID}} &
\colhead{$\log M_*$} &
\colhead{$x$} & \colhead{$y$} & \colhead{$z$} &
\colhead{$v_x$} & \colhead{$v_y$} & \colhead{$v_z$} \\
&
&
\colhead{($\mathrm{M_\odot}$)} &
\colhead{($\mathrm{kpc}$)} &
\colhead{($\mathrm{kpc}$)} &
\colhead{($\mathrm{kpc}$)} &
\colhead{($\kms$)} &
\colhead{($\kms$)} &
\colhead{($\kms$)}
}
\startdata
0 & 44   & 11.2 & 67127 & 76721 & 216007 &  -133 &  -288 &   -76 \\
0 & 109	 & 10.9 & 67078 & 76740 & 216254 &  -437 &   253 &   646 \\
0 & 223  & 10.6 & 67094 & 76742 & 216576 &   138 &   718 &  -519 \\
0 & 313	 & 10.4 & 67078 & 76679 & 216303 &  -293 &  -540 &  1048 \\
0 & 372	 & 10.1 & 67132 & 76633 & 215752 & -1472 & -1495 &  1121 \\
0 & 507	 & 10.1 & 67132 & 76721 & 216100 &  -931 &  -361 &   735 \\
0 & 785  &  9.7 & 67120 & 76609 & 216184 &  -877 &   592 &   818 \\
0 & 1144 & 	9.0 & 67059 & 76716 & 216188 &  1546 & -1526 &  -268 \\
1 & 53	 & 11.1 & 64650 & 72114 & 218315 &    24 &	  -3 & -1260 \\
1 & 57	 & 11.1 & 64727 & 72138 & 217802 &  -632 &  1627 &  -351 \\
1 & 139  & 10.8 & 64705 & 72109 & 217948 &   546 &  1879 &	-623 \\
1 & 387  &	9.8 & 64721 & 71974 & 216370 &	-911 &	-717 &	 499 \\
1 & 765  &	9.7 & 64715 & 72037 & 218071 &  1585 &   978 &	 328 \\
1 & 1146 &	9.4 & 64672 & 71929 & 218099 &	 553 &  -113 &   915 \\
\enddata
\tablecomments{This table is available in its entirety in machine-readable form in the online journal.}
\end{deluxetable*}

\begin{deluxetable*}{ccccccccc}
\label{tab:3dcg_pp}
\tablecaption{Catalog of the PPP CGs in TNG300 at $z=0$.}
\tablehead{
\colhead{\multirow{2}{*}{ID}} & 
\colhead{\multirow{2}{*}{$N_{mem}$}} & 
\colhead{$R_{gr, 3D}$} & 
\colhead{$R_{gr, proj}$} &
\colhead{$\log\rho_{mem}$} & 
\colhead{$\log M_{CG,star}$}& 
\colhead{$\sigma_{3D}$} &
\colhead{$\sigma_{LoS,z}$} & 
\colhead{\multirow{2}{*}{$H_0t_{cr}$}} \\
 &  & 
\colhead{(kpc)} & 
\colhead{(kpc)} & 
\colhead{(Mpc$^{-3}$)} & 
\colhead{($\Mdot$)} & 
\colhead{($\kms$)} & 
\colhead{($\kms$)} & 
}
\startdata
0 & 3 & 29 & 29 & 4.5 & 11.1 & 805 & 412 & 0.002 \\
1 & 3 & 63 & 62 & 3.5 & 10.0 & 921 & 899 & 0.003 \\
2 & 3 & 46 & 45 & 3.6 & 9.9 & 1819 & 845 & 0.001 \\
3 & 4 & 90 & 48 & 3.1 & 11.1 & 1130 & 796 & 0.002 \\
4 & 3 & 49 & 48 & 3.8 & 10.7 & 1654	& 488 &	0.001 \\
\enddata
\tablecomments{This table is available in its entirety in machine-readable form in the online journal.}
\end{deluxetable*}

\begin{deluxetable*}{ccccccccc}
\label{tab:prcg_pp}
\tablecaption{Catalog of the PPV CGs in TNG300 at $z=0$.}
\tablehead{
\colhead{\multirow{2}{*}{ID}} & 
\colhead{\multirow{2}{*}{$N_{mem}$}} & 
\colhead{$R_{gr, 3D}$} & 
\colhead{$R_{gr, proj}$} &
\colhead{$\log\rho_{mem}$} & 
\colhead{$\log M_{CG,star}$}& 
\colhead{$\sigma_{3D}$} &
\colhead{$\sigma_{LoS,z}$} & 
\colhead{\multirow{2}{*}{$H_0t_{cr}$}} \\
 &  & 
\colhead{(kpc)} & 
\colhead{(kpc)} & 
\colhead{(Mpc$^{-3}$)} & 
\colhead{($\Mdot$)} & 
\colhead{($\kms$)} & 
\colhead{($\kms$)} & 
}
\startdata
0 & 8 &  428 & 112 &  1.4 & 11.5 & 1307 & 601 & 0.008 \\
1 & 6 & 1659 & 187 & -0.5 &	11.5 & 1461 & 739 & 0.010 \\
2 & 6 &  889 &  97 &  0.3 & 11.2 & 1334 & 499 & 0.019 \\
3 & 3 &  450 &  70 &  0.9 & 10.7 & 1183 & 298 & 0.015 \\
4 & 3 &  256 &  52 &  1.6 & 11.0 & 2303 & 106 & 0.004 \\
\enddata
\tablecomments{This table is available in its entirety in machine-readable form in the online journal.}
\end{deluxetable*}

\subsection{The Physical Properties of Compact Groups} \label{sec:pp_measurement}

\begin{figure*}
\centering
\includegraphics[width=0.7\linewidth]{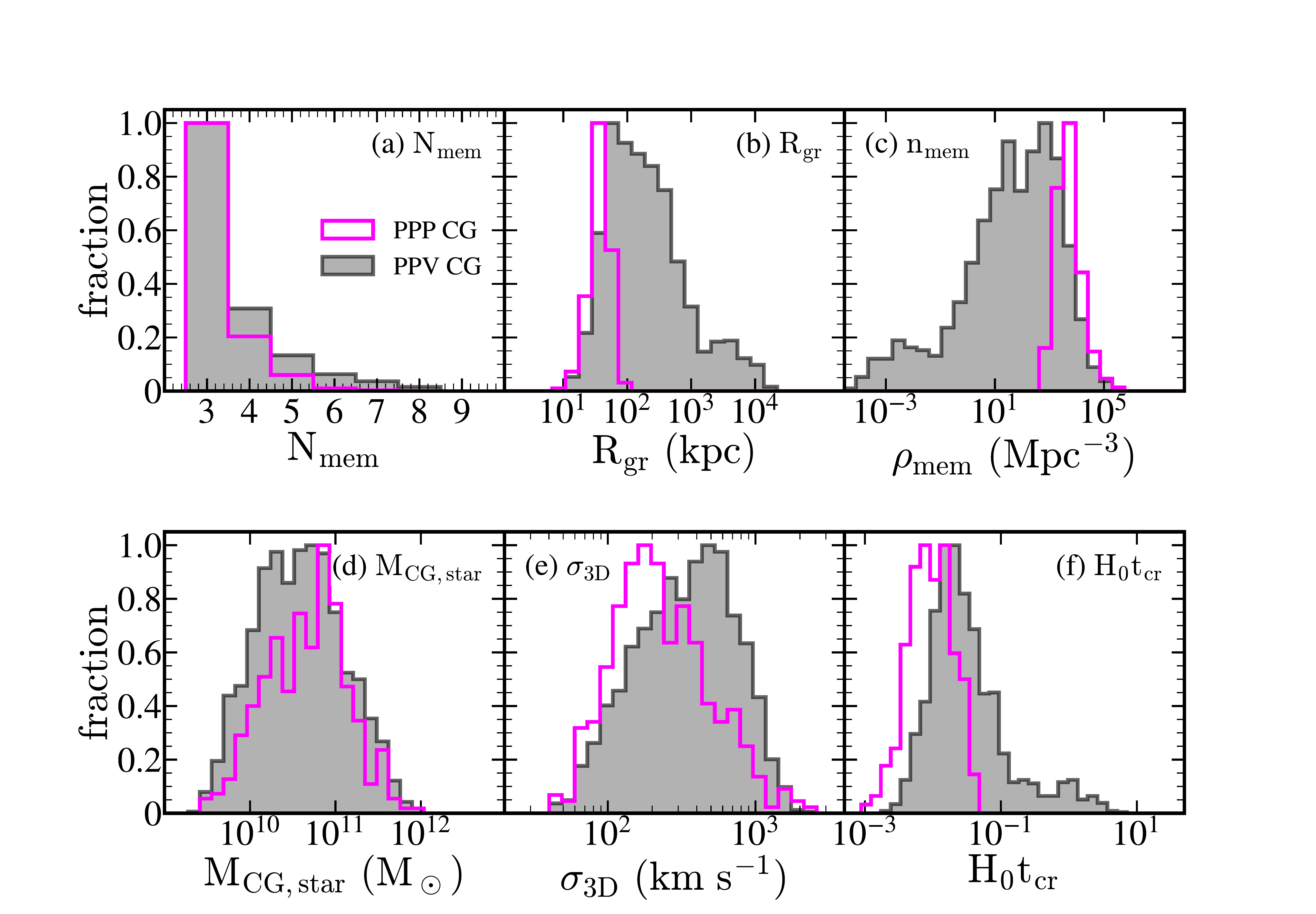}
\caption{Distributions of physical properties of the PPP (the hatched magenta histogram) and PPV (the filled black histogram) CGs: (a) the number of members, (b) 3D size, (c) the number density of member galaxies, (d) the total stellar mass of group members, (e) 3D velocity dispersion of member galaxies, and (f) the dimensionless crossing time. } 
\label{fig:sec4pp_hist}
\end{figure*}

Figure \ref{fig:sec4pp_hist} (a) displays the number of member galaxies in PPP and PPV CGs in TNG300. The majority of PPP ($\sim 78\%$) and PPV ($\sim 64\%$) CGs consist of three-member galaxies. The maximum member number of PPP CGs is nine; 22 PPV systems with ten or more members are excluded.

We next measure the size ($R_{gr,3D}$) of PPP CGs (the magenta histogram in Figure \ref{fig:sec4pp_hist} (b)). To measure the sizes of CGs, we first determine the center of CGs as a center of mass of group determined based on the stellar masses of member galaxies. Then, the maximum distance between the group center and the most distanct member galaxy is the size of PPP CG. The sizes of PPP CGs range from 12 kpc to 164 kpc, with the median size of 57 kpc. In Figure \ref{fig:sec4pp_hist} (b), we also display the sizes of PPV CGs with the black histogram. The typical size of PPV CGs is much larger than the PPP CGs with the median size of 247 kpc within the range of 12 kpc to 24194 kpc. Because a large fraction of PPV CGs are false positives, with member galaxies widely separated along the line-of-sight, their sizes measured in 3D space exhibit a much more extended distribution. The extreme 3D sizes of PPV CGs also indicate that many of these systems are not genuinely `compact' configurations of the type we aim to identify.

Figure \ref{fig:sec4pp_hist} (c) shows the distributions of the number density of CG members ($\rho_{mem}$) of PPP and PPV CGs. Here, $\rho_{mem}$ is defined as the number of member galaxies in each CG divided by the spherical volume determined based on the three-dimensional radius, $R_{gr, 3D}$. The PPP CGs indeed show high number density; the typical number density is $\sim 10^{3.6}\ \rm Mpc^{-3}$. On the contrary, the PPV CGs show a broader distribution of the number density and the typical number density is only $\sim 10^{1.7}\ \rm Mpc^{-3}$. 

In Figure \ref{fig:sec4pp_hist} (d), we plot the group stellar mass distribution of the two types of CGs. The group stellar mass indicates the sum of stellar mass for all group member galaxies. We sum the mass of all stellar particles that are the members of group member subhalos as the group stellar masses. We do not include stellar particles in the intragroup medium. The group stellar mass distributions for the PPP and PPV CGs are comparable because they have a similar number of members with similar masses. The typical group stellar mass is $7.5 \times 10^{10}~\Mdot$ for the PPP CGs and $5.8 \times 10^{10}~\Mdot$ for the PPV CGs. 

One of the important physical properties we use in this work is the velocity dispersion of CGs. We compute the standard deviation of 3D peculiar velocity of CG members as the velocity dispersion of CG (Figure \ref{fig:sec4pp_hist}). Measuring velocity dispersion of small number system is challenging. We thus test various velocity dispersion measurement techniques including the gapper method and the biweight technique \citep{Beers1990}, but the resulting velocity dispersion measurement change little. The velocity dispersion of PPP and PPV CGs range from $45 < \sigma_{3D} (\kms) < 2460$ with the median values of 215 and $350~\kms$, respectively. Again, a large fraction of false positives in PPV CGs lead the velocity dispersion to be overstated comparing to the velocity dispersion of PPP CGs.
 
We show the dimensionless crossing time of our CG samples in Figure \ref{fig:sec4pp_hist} (f). The crossing time of CGs is defined as $t_{cr}=R_{sep}/\sigma_{3D}$ \citep{Hickson1992}, where $R_{sep}$ is the median of three-dimensional separations of all possible pairs of CG members. We multiply by $H_{0}$ to convert the crossing time to the dimensionless crossing time (i.e., $H_{0}t_{cr}$). The typical dimensionless crossing time of the PPP and PPV CG is 0.009 and 0.024, respectively. The large crossing time of PPV CGs results from the apparent large separation between member galaxies. 

\subsection{Velocity Dispersion Function as Group Total Stellar Mass} \label{sec:Msig}

\begin{figure}
\centering
\includegraphics[width=1\linewidth]{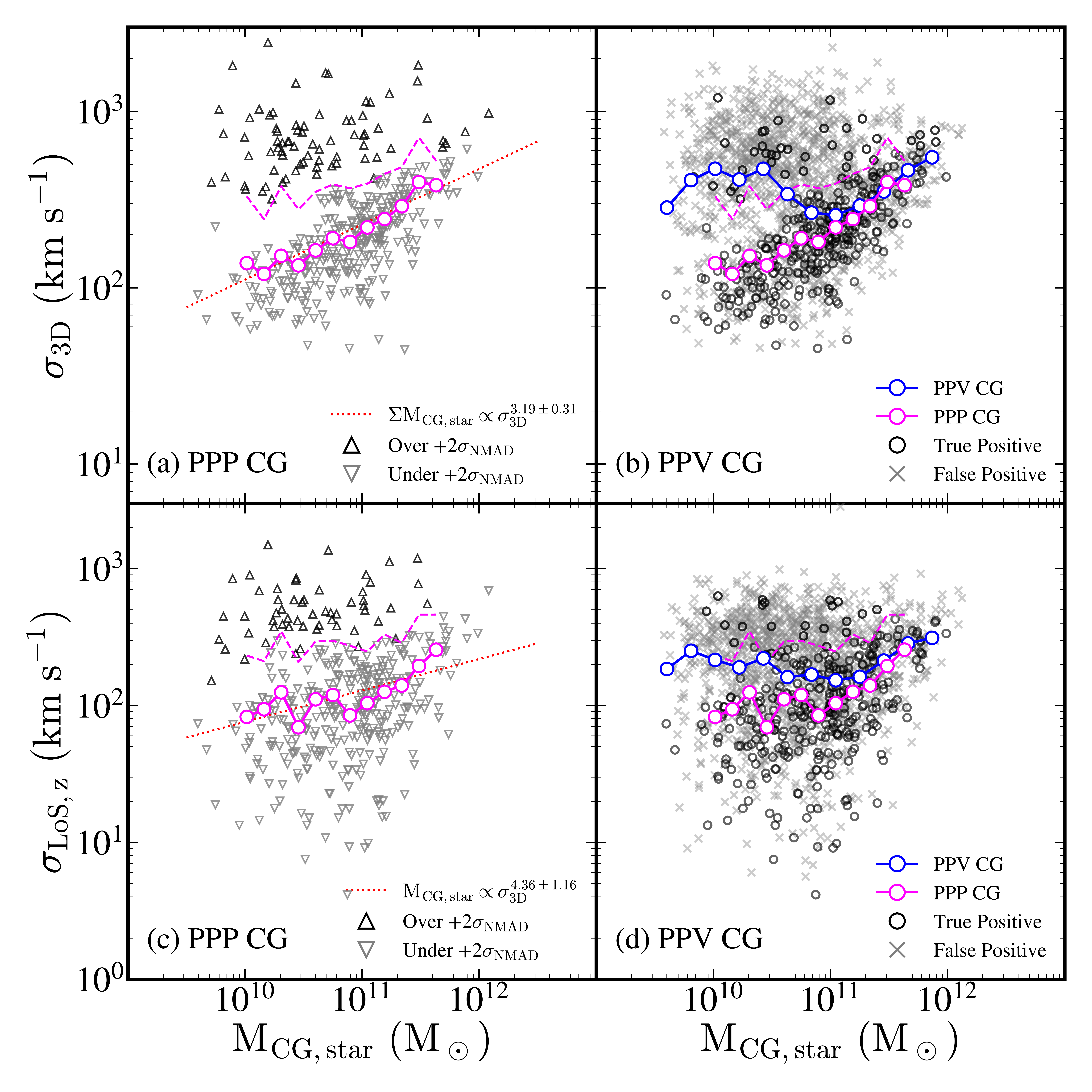}
\caption{(a) Three-dimensional velocity dispersions ($\sigma_{3D}$) as a function of the total stellar mass of the group ($M_{CG,star}$) for the PPP CGs. Magenta circles show median velocity dispersion of each mass bin. The magenta dashed line indicates the boundary located 2$\sigma_{NMAD}$ above the median velocity dispersion in each mass bin separating outliers (upper triangles) from normal CGs (lower triangles). (b) Same as panel (a), but for the PPV CGs. The crosses and circles denote the false and true positives of the PPV CGs respectively. (c) and (d) Same relations as (a) and (b), respectively, but based on the line-of-sight velocity dispersions ($\sigma_{LoS}$) instead of the three-dimensional velocity dispersions. }
\label{fig:3cpr_Msig}
\end{figure}

We investigate the relation between the velocity dispersion (i.e., $\sigma_{3D}$) and the group total stellar mass (i.e., $M_{CG,star}$). Figure \ref{fig:3cpr_Msig} (a) and (b) display the $\sigma_{3D}$ and $M_{CG,star}$ distribution of PPP and PPV CGs. For comparison, we also plot the line-of-sight velocity dispersion (i.e., $\sigma_{LoS,z}$) and $M_{CG,star}$ of PPP and PPV CGs in Figure (c) and (d). We note that the relations shown in the lower right panel of Figure \ref{fig:3cpr_Msig} based on the line-of-sight velocity dispersions can also be derived from the observational CG samples.

In Figure \ref{fig:3cpr_Msig} (a), the majority of PPP CGs show a tight correlation between $\sigma_{3D}$ and $M_{CG,star}$. We compute the median $\sigma_{3D}$ in various $M_{CG,star}$ bins (magenta circles) and derive the best-fit relation (the magenta dotted line): $M_{CG,star}\propto\sigma_{3D}^{3.19\pm0.31}$. The scaling relation of PPP CGs has a similar slope with the scaling relation between the total mass and the velocity dispersions of relaxed galaxy clusters (e.g., \citealp{Evrard2008, Rines2016}) or the stellar mass to stellar velocity dispersion relation of quiescent galaxies (e.g., \citealp{Zahid2016}). The consistent slope of the scaling relation suggests that the majority of CGs are also gravitationally bound systems.

There are some PPP CGs that are offset from the scaling relation with high velocity dispersions (black triangles). We compute the $\sigma_{NMAD}$ in each group stellar mass bin (i.e., the normalized median absolute deviation, \citealp{Stetson1987}): $\sigma_{NMAD} = 1.48 \times (\mathrm{median}(|\sigma_{3D_i}-\mathrm{median}(\sigma_{3D})|)$. We select CGs with a velocity dispersion larger than $2\sigma_{NMAD}$ from the median $\sigma_{3D}$ at given $M_{CG,star}$ as outliers. Among PPP CGs, 76 (20\%) outliers have large velocity dispersions. We will discuss the nature of these outliers in Section \ref{sec:embedded_cg}. 

We derive the same relation for the PPV CGs, as shown in Figure \ref{fig:3cpr_Msig}. We first note that the number of outliers is significantly larger. There are 898 ($\sim 54\%$) systems above the diagnostic boundary we derived based on the PPP CGs (i.e., the $2\sigma_{NMAD}$ outliers), and only 8\% of them are genuine compact systems (true positives). In other words, the boundary in the $M_{CG,star} - \sigma_{3D}$ effectively distinguishes the outliers in the PPV CGs. 

There are 768 ($\sim 46\%$) PPV CGs following the $M_{CG,star} - \sigma_{3D}$ scaling relation of the PPP CGs. However, the consistent scaling relation do not guarantee that they are the similar systems, because 66\% PPV CGs that follow the scaling relation are false positives. The high false positive fraction among the systems showing good correlation indicates that the $M_{CG,star} - \sigma_{3D}$ relation is efficient at removing obvious extreme outliers, but remains imperfect for identifying all false positives.

Figure \ref{fig:3cpr_Msig} (c) and (d) display the similar relations with (a) and (b), but based on the line-of-sight velocity dispersion along with the $z-$axis ($\sigma_{LoS,z}$). We show $\sigma_{LoS,z}$ which is observable quantity unlike $\sigma_{3D}$. 

We compute the median $\sigma_{LoS,z}$ as a function of $M_{CG,star}$. The best-fit relation for PPP CGs is: $M_{CG,star}\propto \sigma_{LoS,z}^{4.36 \pm 1.16}$. Here, the slope is larger than for the $\sigma_{3D} - M_{CG,star}$ relation, but the uncertainty of the slope is also very large. Because $\sigma_{LoS,z}$ is the one-dimensional velocity dispersion is computed based on the $z-$direction velocity, not fully representing the dynamical status as $\sigma_{3D}$, the larger uncertainty in the $\sigma_{LoS,z} - M_{CG,star}$ relation is not surprising. 

There are 56 systems (15\% among PPP CGs) with $\sigma_{LoS}$ larger than $2\sigma_{MAD}$ compared to the median $\sigma_{LoS}$ at given $M_{CG,star}$ Most of these outliers are also outliers when they were selected based on $\sigma_{3D}$ (as in Figure \ref{fig:3cpr_Msig} (a)). The number of outliers among PPV CGs is much larger; 517 PPV CGs (31\%) are outliers with $\sigma_{LoS}$, $2\sigma_{MAD}$ larger than the median $\sigma_{LoS}$ at given $M_{CG,star}$. Among these outliers, only 24 (5\%) systems are true positives, indicating that many of outliers are mere chance alignments, not genuine CGs.

\begin{figure}
\centering
\includegraphics[width=1\linewidth]{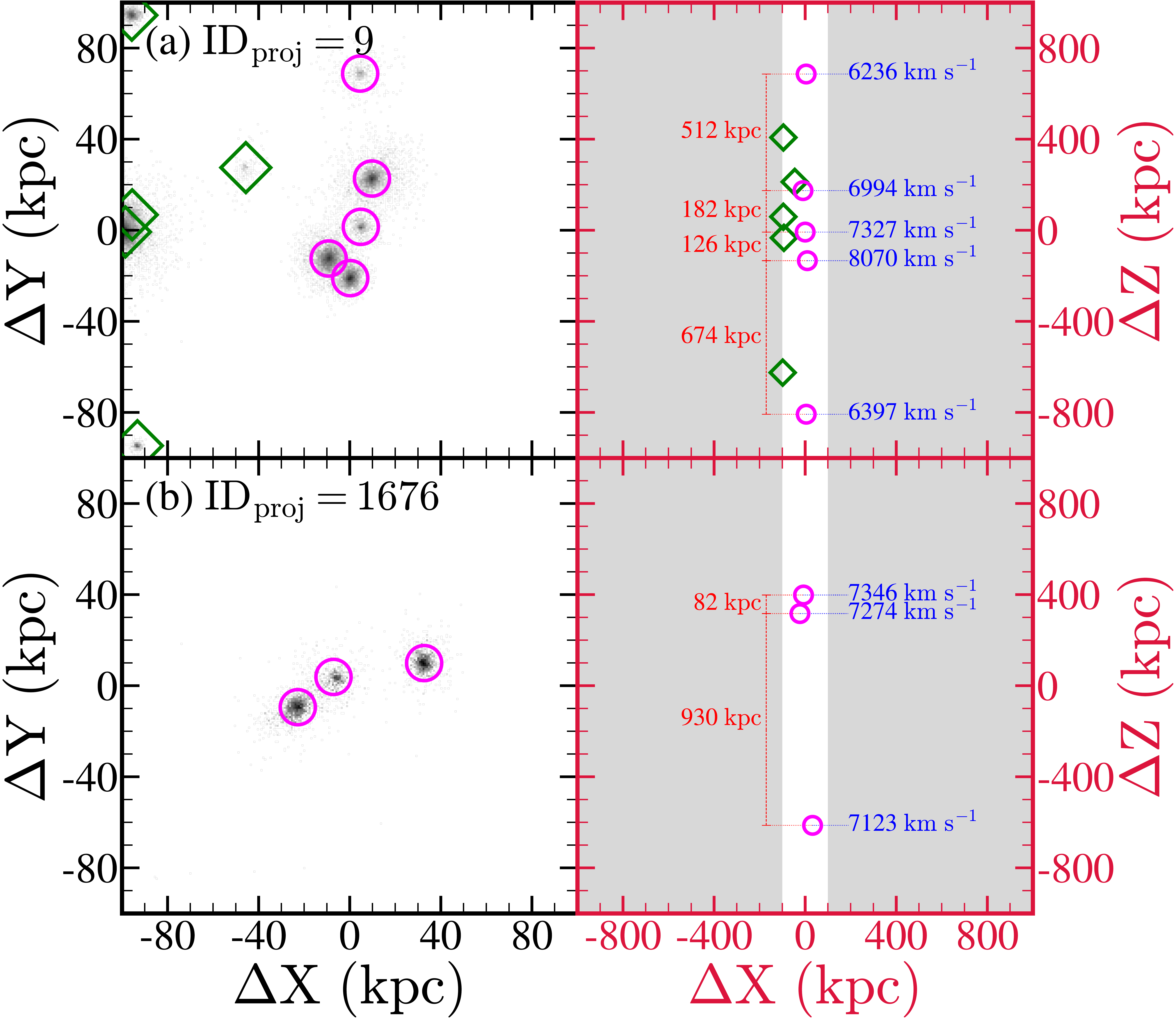}
\caption{Examples of two false positives among PPV CGs. For each example, we display the location of galaxies on the $x\text{-}y$ and $x\text{-}z$ planes, respectively. In each panel, magenta circles indicate the member of PPV CGs and the green diamonds mark non member subhalos.} 
\label{fig:pcg_fp_visuals}
\end{figure}

Figure \ref{fig:pcg_fp_visuals} illustrates representative examples of false positives among PPV CGs. The left panels show the spatial distribution of galaxies in the $x\text{-}y$ plane, while the right panels display the same systems in the $x\text{-}z$ plane. Magenta circles indicate PPV CGs members, and green diamonds mark neighboring subhalos.

In the $x\text{-}y$ plane, the members appear in a compact region. In contrast, the $x\text{-}z$ projections reveal that the physical separations among these galaxies are typically much larger than 73 kpc linking length, indicating that the systems are not physically compact in 3D space. These examples demonstrate that similar line-of-sight velocities do not necessarily imply small physical separations, as galaxies can share coherent peculiar velocities despite being widely separated along the line of sight. Consequently, CGs selected based on the projected and radial linking lengths easily include false positives (i.e., chance alignments).

\subsection{Completeness of Compact Group catalog} \label{sec:subhaloflag}
\balance

We identify CGs using the subhalo catalog constructed by the \texttt{SUBFIND} algorithm, which bundles up particles as a group based on spatial density information. However, not all subhalos identified by the \texttt{SUBFIND} algorithm correspond to galaxies \citep{Nelson2019}. There are many subhalos, particularly among satellite subhalos, that do not have cosmological origins, indicated by \texttt{Subhaloflag} $=0$ in the TNG subhalo catalog. Unlike cosmic origin subhalos that are formed through structure formation and collapse, the non-cosmic origin subhalos are formed through baryonic processes within the already formed galaxies. By the definition, these non-cosmic origin subhalos 1) are not the central subhalo of their host dark matter halo, 2) formed within the virial radius of the parent halo of their host DM halo, and 3) have a DM fraction (i.e., the ratio between the DM mass and the total subhalo mass) lower than 0.8 at the epoch of formation.

In practice, TNG uses the \texttt{SubLink} algorithm to construct merger trees based on both baryonic components (i.e., stars and star-forming gas cells) and dark matter. By comparing the resulting merger trees, subhalos that appear in the baryonic-based tree but contain little or no associated dark matter are classified as non-cosmic origin subhalos (i.e., \texttt{SubhaloFlag} $=0$). In the $z=0$ snapshot, there are 7962 such non-cosmic origin subhalos with $M_{*} > 10^{9}~\Mdot$, corresponding to $\sim 3\%$ of all subhalos in the same stellar mass range. These non-cosmic origin subhalos at $z=0$ typically formed $\sim 4$ Gyr ago.

Excluding the non-cosmic origin subhalos may result in the incompleteness of the CG catalog. CGs are the environment where the frequent galaxy interactions happen due to the extremely high density and the low velocity dispersion of member galaxies. In other words, the member galaxies may undergo frequent galaxy interactions and result in the frequent formation of non-cosmic origin subhalos. If we exclude the non-cosmic origin subhalos, some CGs in an actively interacting phase might be excluded from our identification. 

We thus examine the incompleteness of the CG identification and their impacts on investigation of CG properties by comparing the CG catalogs with/without non-cosmic origins subhalos. For this test, we identify CGs using the same manner described in Section \ref{sec:CGIdentification}, but including all subhalos regardless of their origins. The identification including non-cosmic origin subhalos include 532 PPP CGs and 1853 PPV CGs. By including non-cosmic origin subhalos in the sample, we identify an additional 155 PPP and 247 PPV CGs; however, we also lose 6 PPP and 60 PPV systems due to violations of the mass ratio selection criteria. As a result, the net increase is 149 PPP and 187 PPV systems. We note that the number of PPP CGs increases by $\sim 40\%$, which is significantly higher than the fraction of non-cosmic origin subhalos in the overall subhalo population, indicating that non-cosmic origin subhalos preferentially reside in CG environments.

\begin{figure}
\centering
\includegraphics[width=1\linewidth]{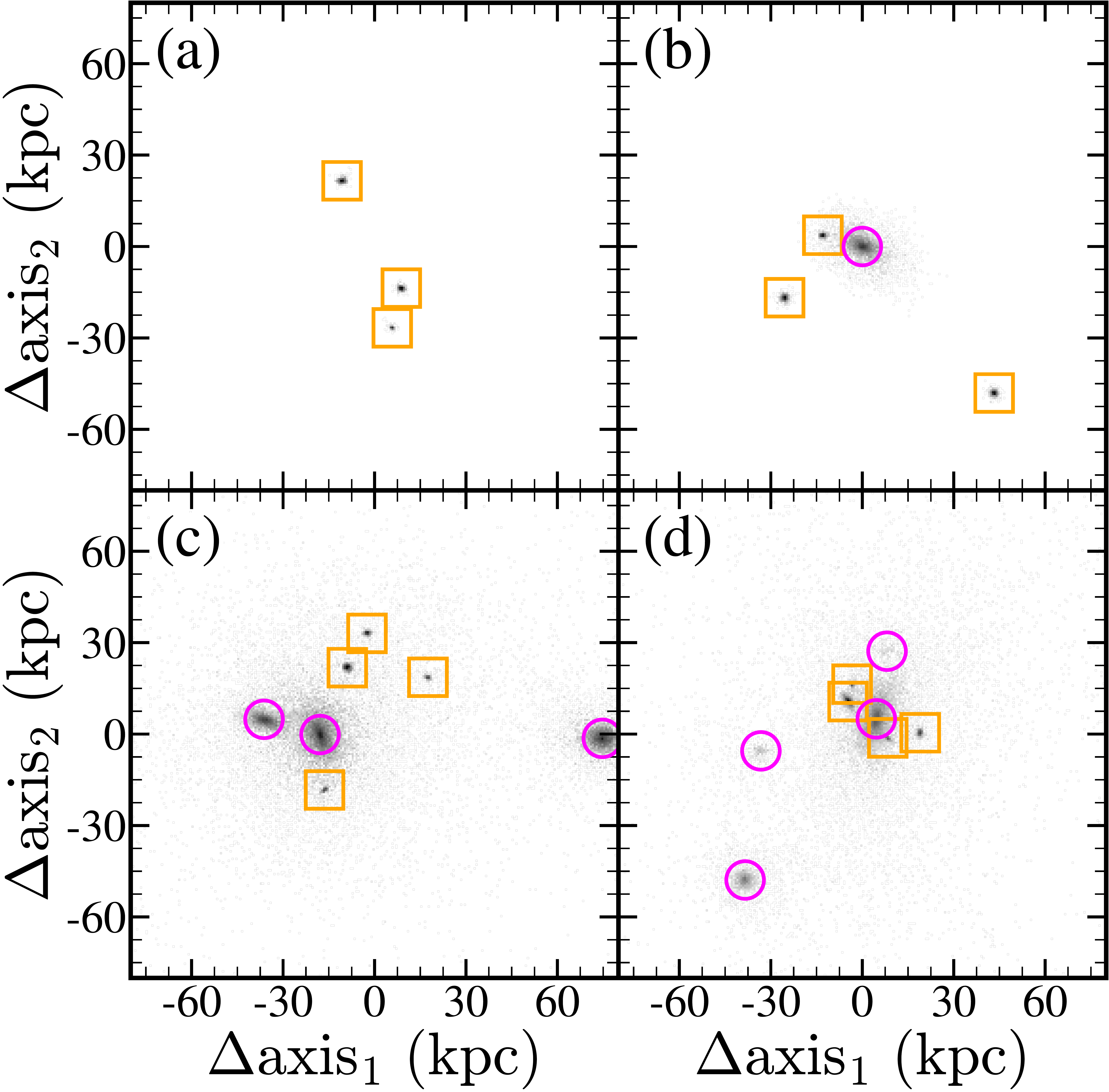}
\caption{Example of CGs identified based on the galaxy sample including non-cosmic origin subhalos. Gray density map shows stellar particle distribution. Orange squares indicate non-cosmic origin subhalos (\texttt{SubhaloFlag} $= 0$) and magenta circles mark cosmic origin subhalos (\texttt{SubhaloFlag} $= 1$).}
\label{fig:Flag_visuals}
\end{figure}

We identify 155 additional systems by including non-cosmic origin subhalos in the galaxy sample used for CG identification. These newly identified systems fall into four categories. First, eight systems are composed entirely of non-cosmic-origin subhalos; Figure \ref{fig:Flag_visuals} (a) shows an example. In this plot, the gray density map indicates the number density of stellar particles, and the yellow squares mark the positions of the non-cosmic origin subhalos. Second, 129 systems are identified as CGs only after the inclusion of one or more non-cosmic origin subhalos alongside one or two cosmic origin subhalos (Figure \ref{fig:Flag_visuals} (b)). Third, 12 systems are classified as CGs because non-cosmic origin subhalos act as bridges between widely separated cosmic origin subhalos, linking them into a single system. Finally, six systems are classified as PPP FoF groups but were not initially identified as PPP CGs due to the large mass contrasts among their member galaxies; the inclusion of non-cosmic origin subhalos reduces these contrasts, enabling their identification as CGs (Figure \ref{fig:Flag_visuals} (d)).

CGs composed exclusively of non-cosmic-origin subhalos are particularly intriguing, and we investigate their nature in more detail. These systems are typically embedded within larger halos. Of the eight systems, six reside within $R_{200}$ of a massive host halo with $M_{\mathrm{halo}} > 10^{13}~\Mdot$ (based on the FoF halo mass), while the remaining two lie outside $R_{200}$ at $z=0$. The individual member subhalos are formed as satellites of a common host halo at different epochs and subsequently assemble into compact configurations as they orbit within the host halo.

\begin{figure*}
\centering
\includegraphics[width=0.9\linewidth]{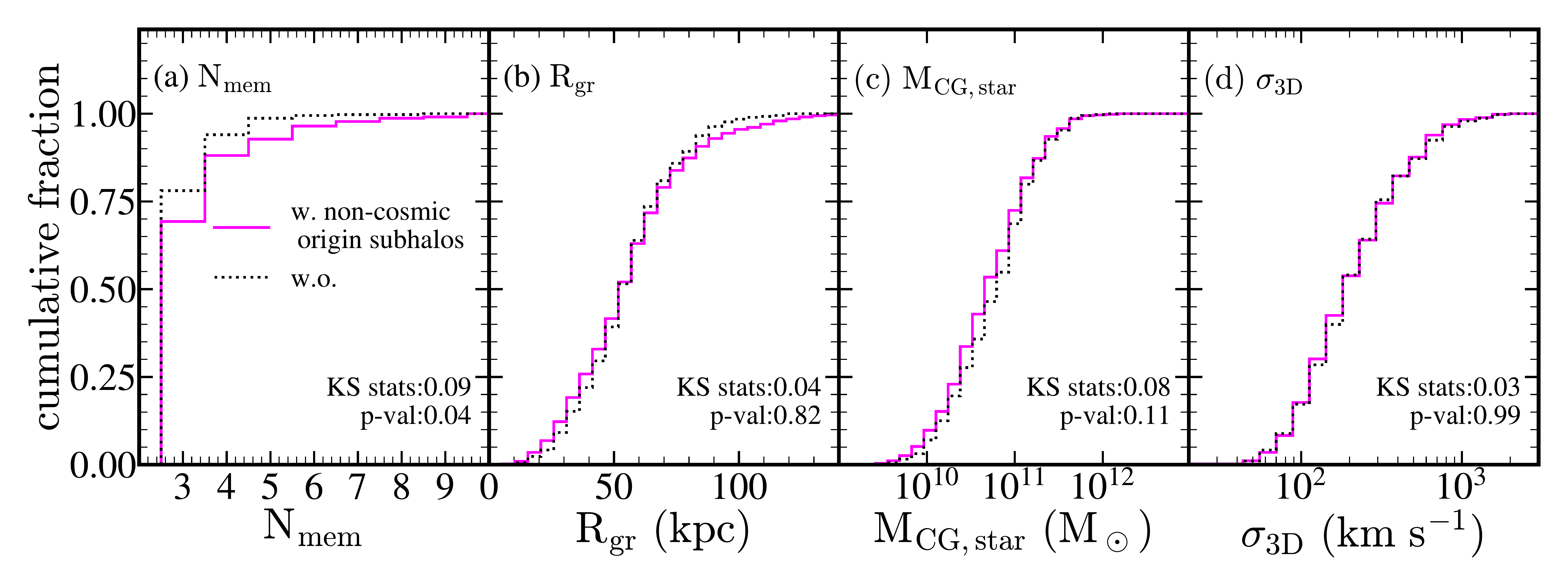}
\caption{Comparison between CGs identified based only on cosmic origin subhalos (black dotted lines) and those identified with non-cosmic origins subhalos (magenta solid lines). We compare the cumulative distributions of (a) the number of members, (b) group size, (c) the total stellar mass of groups ($\Sigma_{*}$), and (d) three dimensional velocity dispersion of member galaxies. The KS test probability and the $p-$value are listed in each panel.}
\label{fig:flag_ppcomp}
\end{figure*}

We examine the physical properties of newly added CGs to examine the impact of inclusion of non-cosmic origin subhalos. Figure \ref{fig:flag_ppcomp} compares the cumulative distributions of (a) the number of members, (b) the 3D size of groups, (c) the total group stellar mass, and (d) the three-dimensional velocity dispersions of the original and the new CG catalogs. We apply the Kolmogorov-Smirnov (KS) tests. The KS test indicates that the two CG systems are basically drawn from the same parental distributions (with the p-values over 0.05). In other words, the inclusion of non-cosmic origin subhalos would not affect any investigation on the physical properties of CGs we identify based on simulations. 

We next investigate the impact of inclusion of non-cosmic origin subhalos in determination of CG abundance. The abundance of CG is often examined to understand the evolution of CGs (e.g., \citealp{MendesdeOliveira1991, Barton1996, Mendel2011, Pompei2012, Sohn2015, Sohn2016, Hartsuiker2020}). If CGs are short-lived structure due to active mergers between member galaxies, the number density of CGs may decrease over the time, unless they are newly formed over the time. Investigating the CG abundance is obviously subject to the CG identification. For example, using the CG sample excluding CGs in an active interacting phase would bias the CG abundance estimates. In other words, excluding non-cosmic origin subhalos in CG identification may result in a bias in CG abundance estimates. 

\begin{figure}
\centering
\includegraphics[width=0.95\linewidth]{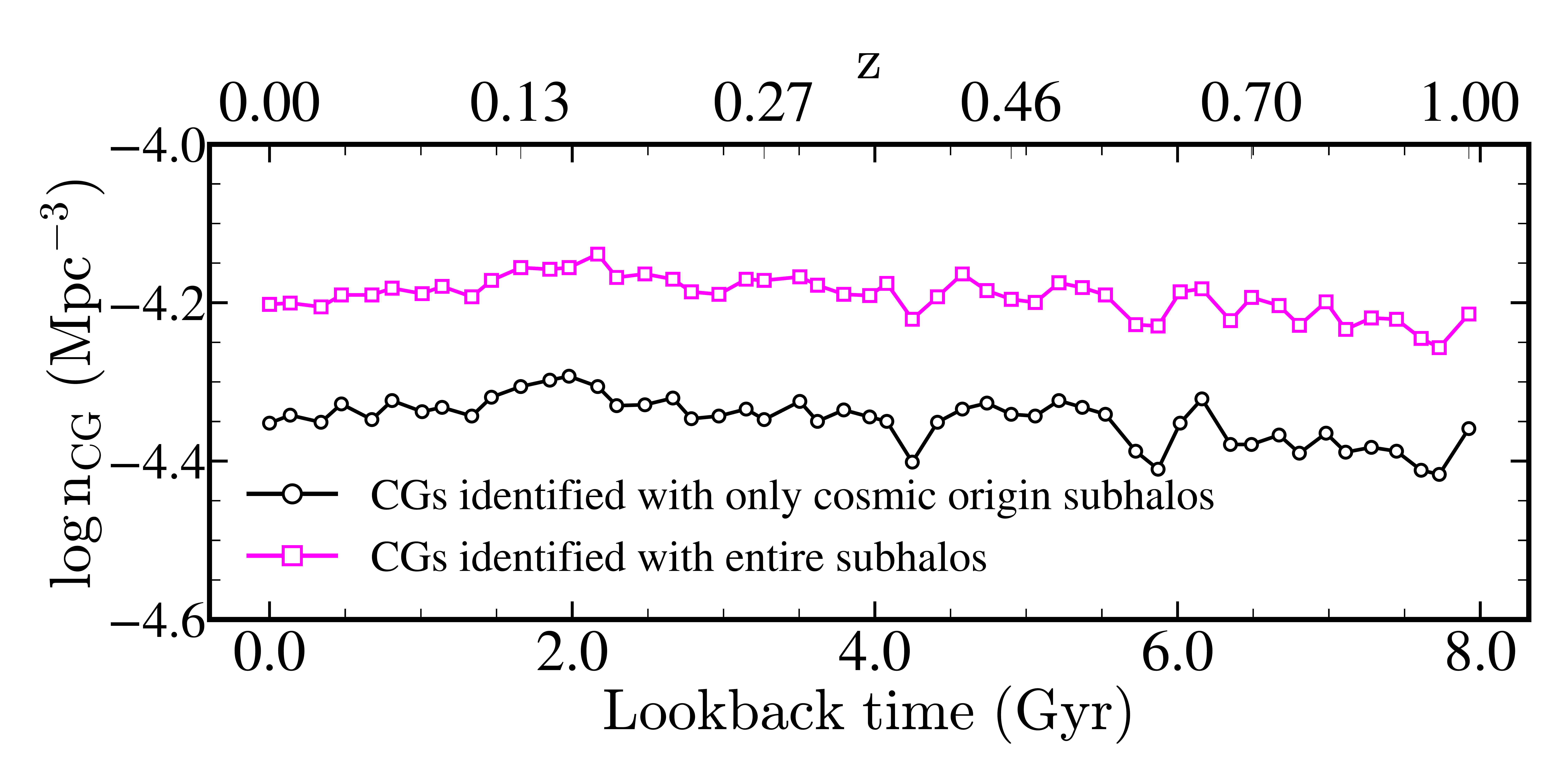}
\caption{Number density of TNG CGs as a function of redshift (lookback time). Magenta squares and black circles indicate the number density of CGs with/without non-cosmic origin subhalos.} 
\label{fig:flag_CGnumdens}
\end{figure}

We thus investigate the potential bias introduced by exclusion of non-cosmic origin subhalos in CG abundance. We identify PPP CGs based on the same technique we applied to $z = 0$ snapshot at higher redshift snapshots, i.e., $z < 1$. We use two subhalo samples including/excluding non-cosmic origin subhalos. Figure \ref{fig:flag_CGnumdens} displays the number density of CGs as a function of redshift: magenta squares for CGs identified with non-cosmic origin subhalos and black circles for CGs identified only with cosmic origin subhalos.

The abundance of CGs not only varies little as a function of redshift in both cases. The CGs including non-cosmic origin subhalos are clearly more abundant than the CGs identified without non-cosmic origin subhalos. However, in both cases, the number density remains constant at $z < 1$, indicating that 1) CGs do not disappear in short time scale (e.g., \citealp{Athanassoula1997, Governato1991}) or 2) CGs do disappear due to mergers between members but new CGs are replenished with similar rate soon (e.g., \citealp{Diaferio1994}). We will investigate the redshift evolution of CGs and their abundances in our forthcoming paper. 

The more important implication of Figure \ref{fig:flag_CGnumdens} is that the redshift evolution of both CG samples shows an essentially identical trend. The consistent trend indicates that the exclusion of non-cosmic origin subhalos only affects to the absolute count of CGs, and the number of missing CGs based on the cosmic origin subhalos is fairly consistent at $z < 1$. Thus, inclusion of non-cosmic origin subhalos in CG identification has little impact on tracing the evolution of CG number density.

\section{Comparison between TNG and SDSS Compact Groups} \label{sec:CG_Obs_comparison}

We compare TNG CGs with the CGs identified based on spectroscopic observations. We particularly use the CGs identified based on SDSS DR12 in \citet{Sohn2016}. These SDSS CGs are identified based on the projected and radial FoF algorithm with $\Delta R_{proj} = 50~\hh$ kpc and with $|\Delta cz| < 1000~\kms$, similar to our PPV CGs in TNG simulations. Because these CGs do not rely on Hickson criteria unlike other observed CGs, the direct comparison with TNG CGs are relatively straightforward.

\citet{Sohn2016} published three CG catalogs based on SDSS spectroscopy. They used three different galaxy samples; their MLCG catalog is based on the magnitude-limited sample with $r < 17.77$ and their V1CG and V2CG catalogs are based on the volume-limited samples with $M_{r} = -19$ mag and within $0.01 < z < 0.0741$ (V1) and with $M_{r} < -20$ and within $0.01 < z < 0.1154$ (V2), respectively. We use the V1CGs for comparing with TNG CGs because V1CGs are selected based the volume limited sample similar to our TNG stellar mass limited subhalo sample.

For fair comparison with the observed CGs, we modify our TNG CG identification by imposing the stellar mass limit corresponding on the absolute magnitude limit applied to the V1CG sample. \citet{Sohn2016} provide both $r-$band absolute magnitude and the galaxy stellar mass estimated from LePHARE spectral energy distribution fitting \citep{Arnouts1999, Ilbert2006}. Based on these measurements, we derive the empirical relation between the two properties: $\log (M_{*} /\Mdot) = A M_{r} + B$ with $A=-0.52\pm0.01$ and $B=-0.37\pm0.21$. The absolute magnitude limit (i.e., $M_{r} = -19$) of the V1CG galaxy sample corresponds to the stellar mass limit of $\log (M_{*} / \Mdot) = 9.60$. We thus re-identify 229 PPP CGs and 897 PPV CGs by applying the FoF algorithm to cosmic origin subhalos with $\log (M_{*} / \Mdot) \geq 9.60$ in TNG300. Hereafter, we refer to these samples as V1 PPP CGs and V1 PPV CGs, respectively.

\begin{figure*}
\centering
\includegraphics[width=0.9\linewidth]{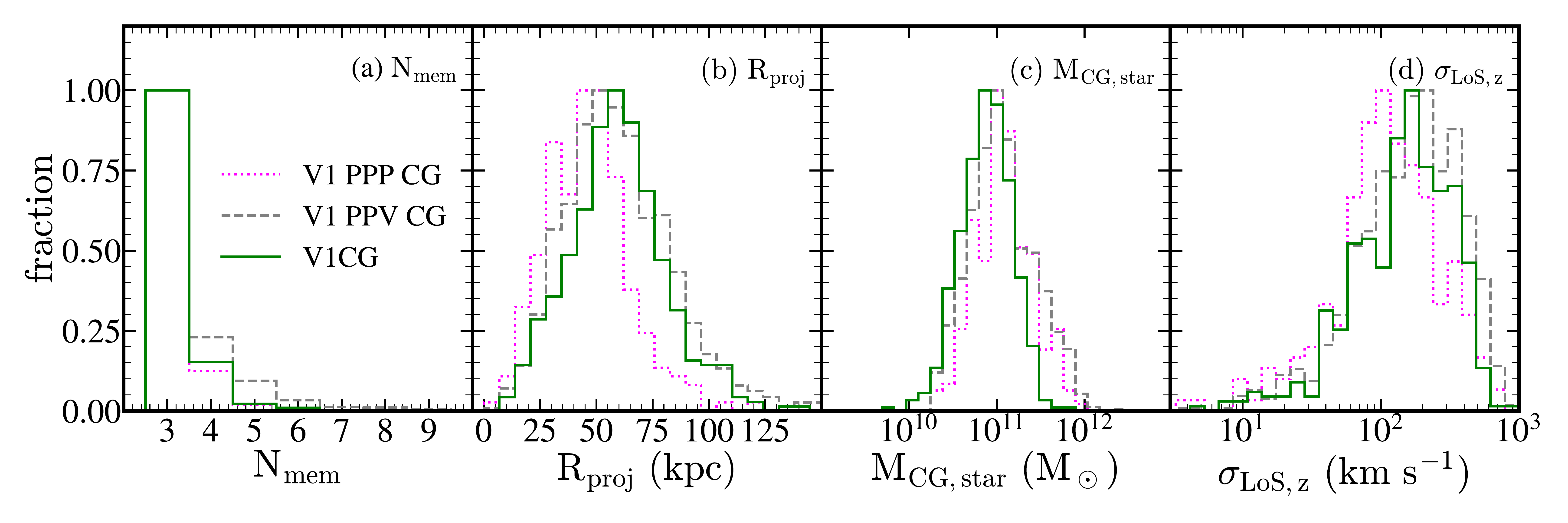}
\caption{Distributions of physical properties of the V1 PPP CGs (the dotted magenta line), V1 PPV CGs (the dashed gray line), and V1CGs (solid green line), including (a) the number of members, (b) the projected size, (c) the total stellar mass of group members, and (d) the line-of-sight velocity dispersion of member galaxies. Here, V1 PPP CGs and V1 PPV CGs are systems identified based on the stellar mass limited sample, mocking the observed volume-limited sample.}
\label{fig:MLCGs_pphist}
\end{figure*}

Figure \ref{fig:MLCGs_pphist} (a) compares the distribution of number of member galaxies of TNG and observed CGs. In terms of number of members, the observed and TNG CGs are indistinguishable. We note that a larger fraction of V1 PPV CGs consist of $N_{mem} \geq 4$ compared to the observed CGs. 

We also compare the normalized distributions of group projected size in panel (b) in Figure \ref{fig:MLCGs_pphist}. The size of V1 PPV and observed CGs are comparable. The high $p-$value (0.20) from the KS test indicates that the null hypothesis that the two size distributions are derived from the same parental distribution cannot be ruled out. In contrast, the size distributions of V1 PPP and observed CGs differ significantly. The median projected size of the V1 PPP CGs (46 kpc) is smaller the median size (59 kpc) of observed V1CGs. The observed V1CGs include larger CGs that the V1 PPP CGs; the low $p-$value ($3.4 \times 10^{-15}$) from KS test suggests that the two size distributions are not derived from the same parental distributions. 

The difference between the three CG catalogs is most distinctive in the  $M_{CG,star}$ distribution (Figure \ref{fig:MLCGs_pphist} (c)). The median $M_{CG,star}$ for the V1 PPP CGs, V1 PPV CGs and V1CGs are $10^{11.1}\ \Mdot$, $10^{11.0}\ \Mdot$, $10^{10.9}\ \Mdot$, respectively. In general, the simulated CGs tend to include slightly more CGs with higher $M_{CG,star}$ and less CGs with lower $M_{CG,star}$. Indeed, the low $p-$values ($< 8.0 \times 10^{-13}$) of the KS tests between the observed and simulated CGs suggest that the $M_{CG,star}$ of the observed and simulated CGs are not derived from the same parental distributions. The differences in the $M_{CG,star}$ range may result from different galaxy sample selections; the observed galaxy sample is constructed based on the absolute magnitude limit and the simulated galaxy sample is built based on the magnitude converted from the stellar mass. 

Figure \ref{fig:MLCGs_pphist} (d) compares the line-of-sight velocity dispersion of three types of CGs. The median velocity dispersion for the V1 PPP CGs, V1 PPV CGs and V1CGs are 110 $\kms$, 172 $\kms$, 161 $\kms$, respectively. The V1 PPP CGs tend to have lower velocity dispersions compared to other CGs. Because both V1 PPV CGs and observed CGs are identified based on the generous radial velocity difference among group members ($|\Delta V| < 1000~\kms$), the larger velocity dispersions of these CGs are not surprising. 

\begin{figure*}
\centering
\includegraphics[width=0.9\linewidth]{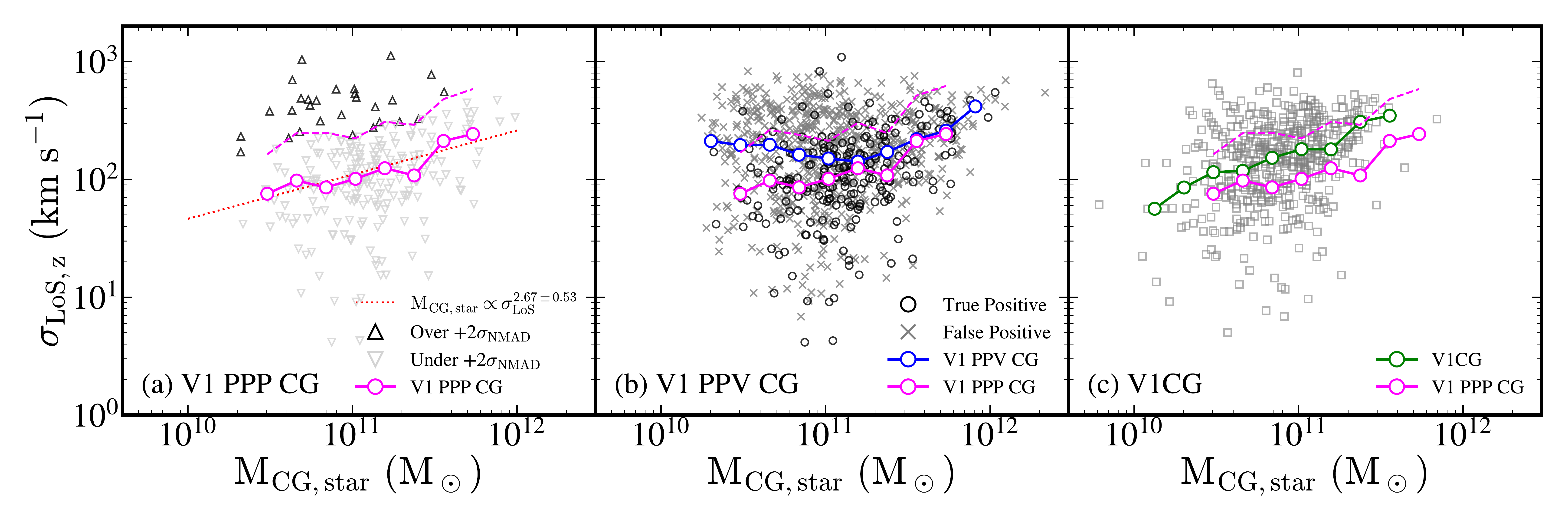}
\caption{(a) Line-of-sight velocity dispersion as a function of the group stellar mass for V1 PPP CGs. The magenta circles indicate the median line-of-sight velocity dispersion at various group stellar mass bins. The magenta dashed line shows the $2\sigma_{NMAD}$ boundary, separating normal CGs (lower triangles) and outliers (upper triangles). (b) Same as (a), but for V1 PPV CGs. Blue circles indicate the similar median $\sigma_{LoS}$s as a function of group stellar mass. We mark true (black circles) and false (gray crosses) positives that are separated based on the physical concentrations. (c) Same as (a), but for the observed V1CGs. Green circles indicate the median $\sigma_{LoS}$s at each group stellar mass bins. The median $\sigma_{LoS}$ of observed CGs is significantly larger than V1 PPP CGs. } 
\label{fig:MLCGs_Msig}
\end{figure*}

Figure \ref{fig:MLCGs_Msig} displays the line-of-sight velocity dispersion as a function of the total group stellar mass for the V1 PPP CGs, V1 PPV CGs and V1CGs. We estimate best-fit relation from the median $\sigma_{LoS}$ in various $M_{CG,star}$ bins: $M_{CG,star}\propto\sigma_{LoS}^{2.67\pm0.53}$. Among the V1 PPP CGs, 22 systems (10\%) lie more than $2\sigma_{\mathrm{NMAD}}$ above this scaling relation. 

Figure \ref{fig:MLCGs_Msig} (b) shows the corresponding results for the V1 PPV CGs. In this sample, 24\% systems (213) are identified as outliers above the $2\sigma_{\mathrm{NMAD}}$ boundary. While 24\% (218) of the entire V1 PPV CG sample are true positives, only 5\% systems (11) among the outliers correspond to genuine V1 PPP CGs. This demonstrates that the outlier region is predominantly populated by false positives, confirming that the conclusion drawn in Section \ref{sec:Msig} remains valid even after applying a mass limit consistent with the observed samples.

Figure \ref{fig:MLCGs_Msig} (c) shows that 23 \% (111) of V1CGs are outliers above the 2$\sigma_{NMAD}$ of scaling relation between $M_{CG,star}$ and $\sigma_{LoS}$. The large outlier fraction suggests that the V1CG sample include a large fraction of chance alignments.

\section{Environment of Compact Groups} \label{sec:embedded_cg}

Identifying embedded CGs is useful to understand the environment dependence in physical property of CGs. In this regard, many previous studies estimate the fraction of embedded CGs. For example, \citet{Ramella1994} found that 76\% of 38 HCGs reside in rich loose groups, whereas \citet{DiazGimenez2015} reported that only 27\% of their 2MASS CGs are embedded. The wide range of fraction of embedded CG reflects differences in CG identification scheme and the definition of the environments and the embeddedness. \citet{Ramella1994} defined embedded CG based on the number of neighbors within a comoving cylinder of projected radius 1.5 Mpc, whereas \citet{DiazGimenez2015} determined embedded CGs by cross-matching with galaxy-group identified by the FoF algorithm. 

\begin{figure}
\centering
\includegraphics[width=1\linewidth]{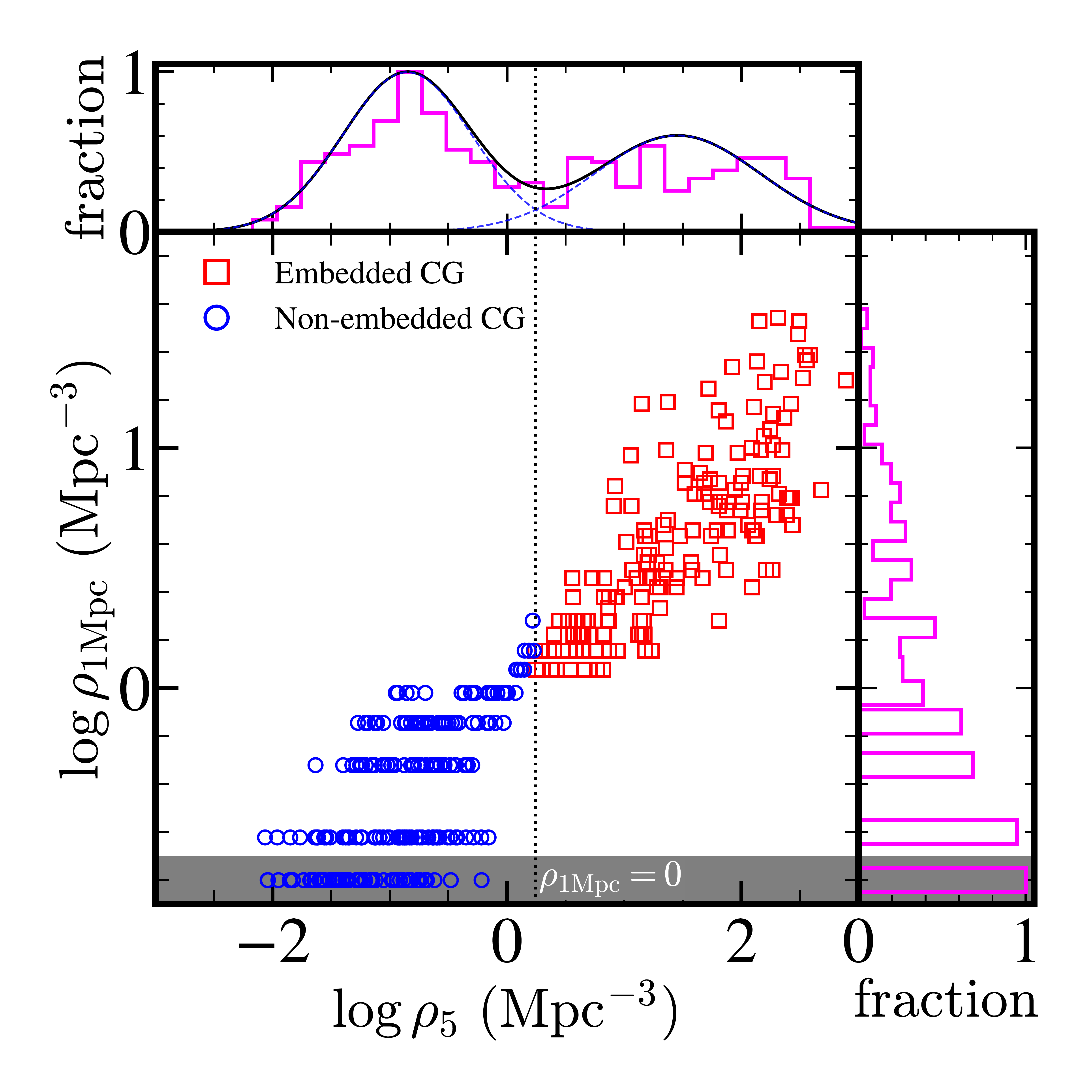}
\caption{Distribution of environment density for PPP CGs defined in two different methods. The magenta histograms and square, circle markers indicate the density distribution of PPP CGs. Dotted line indicates the boundary for embedded (red squares) and non-embedded (blue circles) CGs.}
\label{fig:environment}
\end{figure}

We examine environments of PPP CGs based on two galaxy density estimators. First, we define the local density, $\rho_{5}$, as the number density of galaxies within a spherical volume that encloses the fifth-nearest neighbor:
\begin{equation}
\rho_{5} = \frac{5}{(4\pi R_{\mathrm{3D, 5th}}^{3})/3}.
\end{equation}
When identifying neighboring galaxies, we exclude the CG member galaxies. We then compute $\rho_{\mathrm{1~Mpc}}$, defined as the number density of galaxies within a fixed spherical volume of radius 1 Mpc, again excluding the CG members. These two density estimators together characterize the environments of CGs on both local and larger (i.e., within 1 Mpc) scales.

Figure \ref{fig:environment} shows the distribution of $\rho_{5}$ versus $\rho_{\mathrm{1~Mpc}}$ for PPP CGs. Overall, $\rho_{5}$ shows a positive correlation with $\rho_{\mathrm{1~Mpc}}$, as expected. The scatter in this relation suggests that some CGs reside in locally dense regions embedded within globally low-density environments, or are locally isolated despite being located in larger-scale overdense structures. In addition, we identify 57 extremely isolated systems (indicated by the shaded region in Figure \ref{fig:environment}) that have no neighboring galaxies within 1 Mpc.

We apply a Gaussian mixture model to separate the underlying populations in the $\rho_{5}$ distribution. The dashed lines in the upper panel of Figure~\ref{fig:environment} show the two best-fit Gaussian components that describe the overall $\rho_{5}$ distribution. The two components have mean values of $\mu_1 = -0.9$ and $\mu_2 = 1.5$ in $\log\rho_5$, with standard deviations of $\sigma_1 = 0.6$ and $\sigma_2 = 0.7$, respectively. To test the bimodality in distribution of $\rho_{5}$, we use Ashman's D statistic \citep{Ashman1994}:
\begin{equation}
    D = \frac{|\mu_1-\mu_2|}{\sqrt{2(\sigma_1^2+\sigma_2^2)}}
\end{equation}
The resulting value is $D=3.6$, exceeding the conventional threshold of $D=2$, indicating a statistically significant bimodality in the $\rho_{5}$ distribution. The two Gaussian components intersect at $\log \rho_{5} \sim 1.7$ Mpc$^{-3}$, which we adopt as the boundary separating low and high density environments. Based on this criterion, 44\% of systems are within a high density environment with $\log \rho_{5} > 1.7$ Mpc$^{-3}$. We refer to these systems as the embedded CGs. 

\begin{figure*}
\centering
\includegraphics[width=0.7\linewidth]{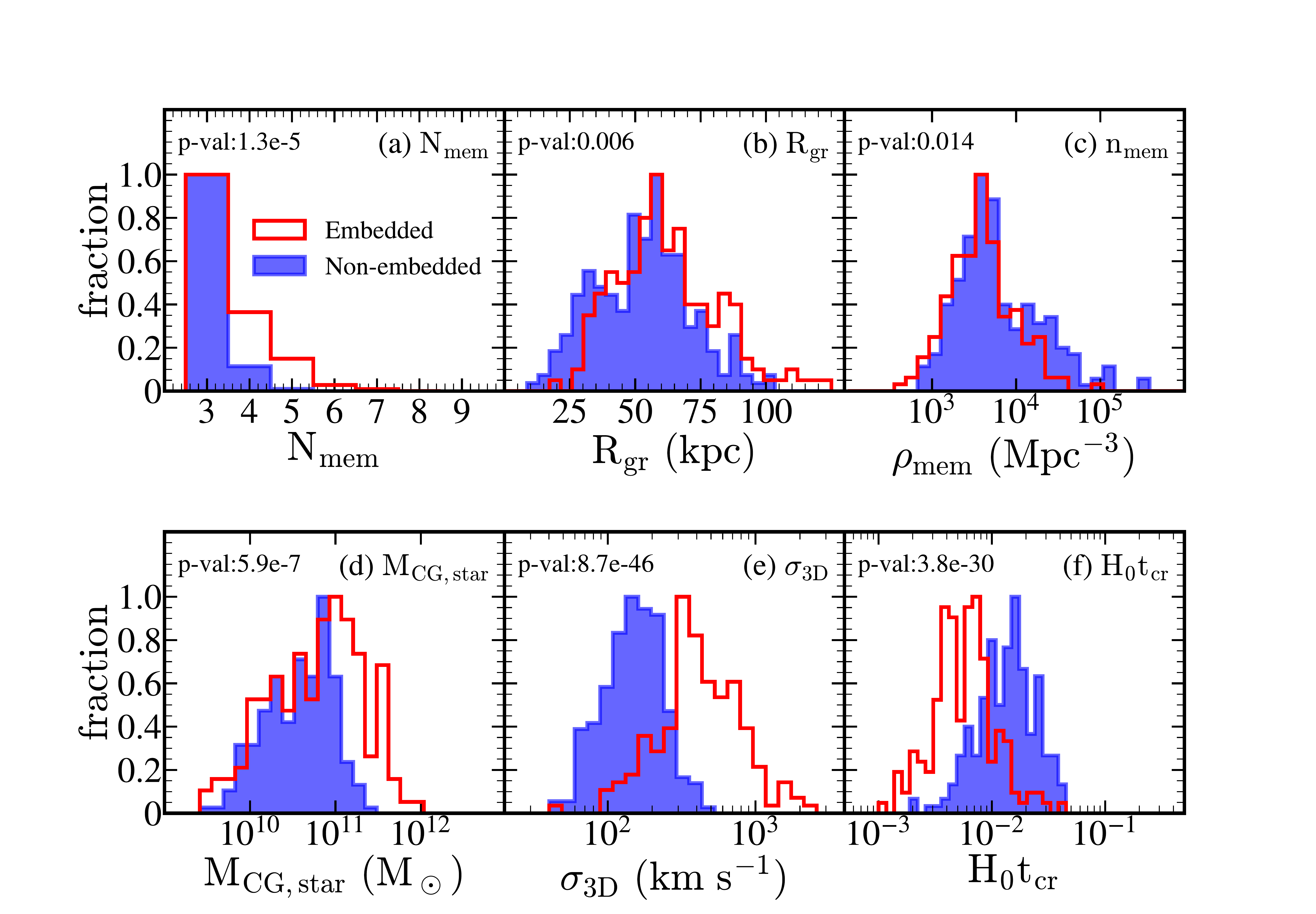}
\caption{Distributions of physical properties of embedded (open red histogram) and non-embedded (filled blue histogram) CGs: (a) the number of members, (b) 3D size, (c) the number density of member galaxies, (d) the total stellar mass of group members, (e) 3D velocity dispersion of member galaxies, and (f) the dimensionless crossing time. }
\label{fig:env_pp}
\end{figure*}

Figure \ref{fig:env_pp} compares the physical properties of embedded and non-embedded CGs in less dense environments including (a) number of members, (b) group size, (c) the member number density of group, (d) group total stellar mass, (e) the line-of-sight velocity dispersion of member galaxies, and (f) the dimensionless crossing time. Interestingly, the two CG populations show distinctive distributions in all parameters we explore. We employ the KS test to determine if the physical properties of embedded and non-embedded CGs are drawn from the same underlying population. For all properties displayed in Figure 11, the resulting $p$-values are consistently lower than 0.05, indicating that the two CG populations are statistically distinct.

\begin{figure}
\centering
\includegraphics[width=1\linewidth]{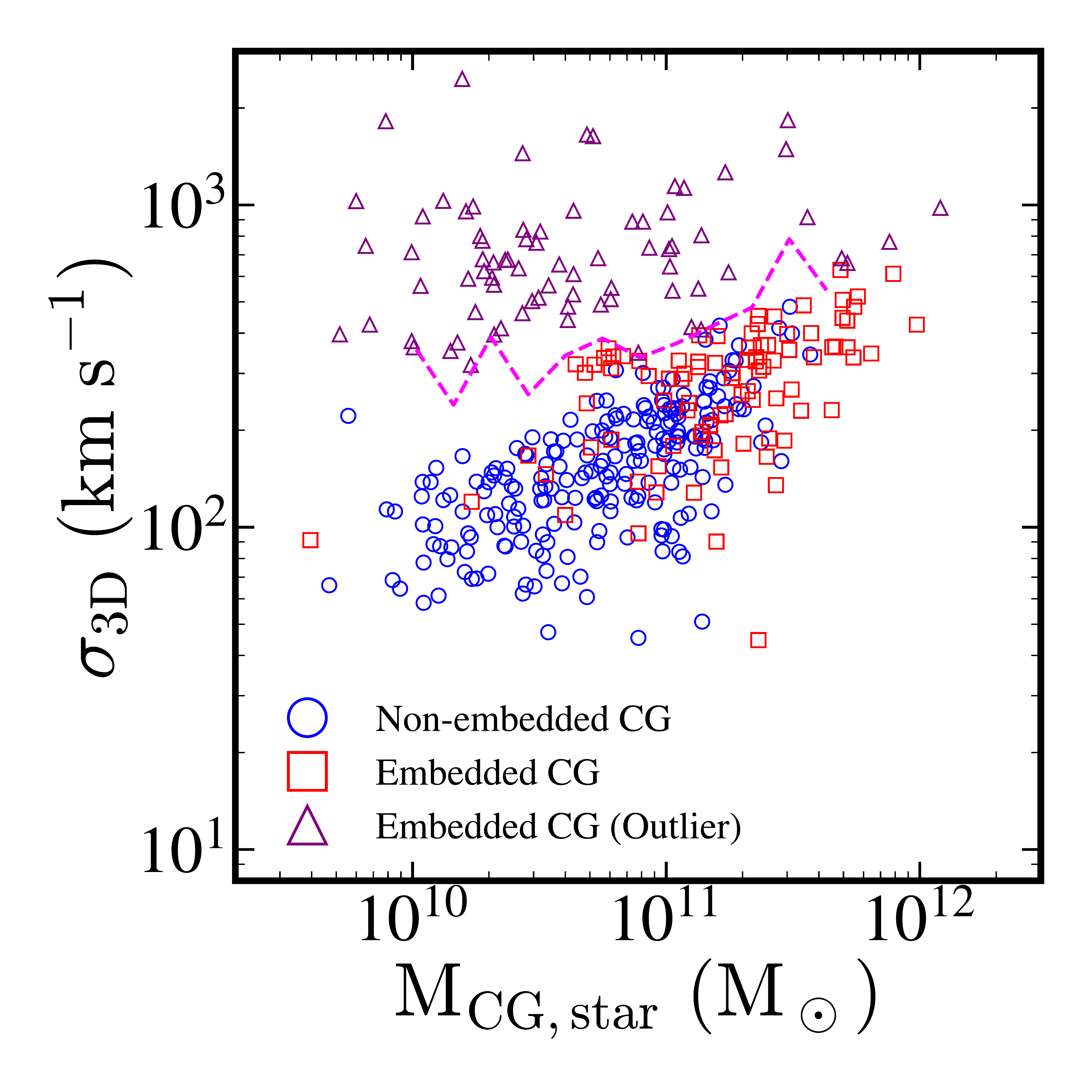}
\caption{(a) Velocity dispersion as a function of the group stellar mass for PPP CGs. The blue circle indicate non-embedded CGs. The red square and purple triangle indicate embedded CGs following and diverging $M_{CG,star}-\sigma_{3D}$ scaling relation respectively. (b) Distribution of environment density for PPP CGs defined in two different methods.}
\label{fig:env_Msig}
\end{figure}

Figure \ref{fig:env_Msig} (a) displays the three dimensional velocity dispersion as a function of group total stellar mass for embedded (red squares) and non-embedded (blue circles) CGs. Non-embedded CGs generally follow $M_{CG,star}-\sigma_{3D}$ scaling relation in Figure \ref{fig:3cpr_Msig}. A large fraction of embedded CGs also follow the scaling relation, and they generally have higher group stellar mass and velocity dispersion compared to non-embedded CGs (as expected from Figure \ref{fig:env_pp} (d) and (e)). 

In Figure \ref{fig:env_Msig} (a), there are PPP CGs (purple triangles) located in the region where the fraction of false positives is high in the PPV CGs. These outlier systems are all concentrated in the three-dimensional space, but they show much higher velocity dispersion than the expected values from given group total stellar mass. Interestingly, these systems are all embedded CGs (Figure \ref{fig:env_Msig} (b)). Most of these systems are parts of massive galaxy clusters. Because they are not relaxed independent systems and their velocity dispersions are inflated by the large potential in environment, the total stellar mass and velocity dispersion relations are offset from the other CGs. 

\begin{figure}
\centering
\includegraphics[width=1\linewidth]{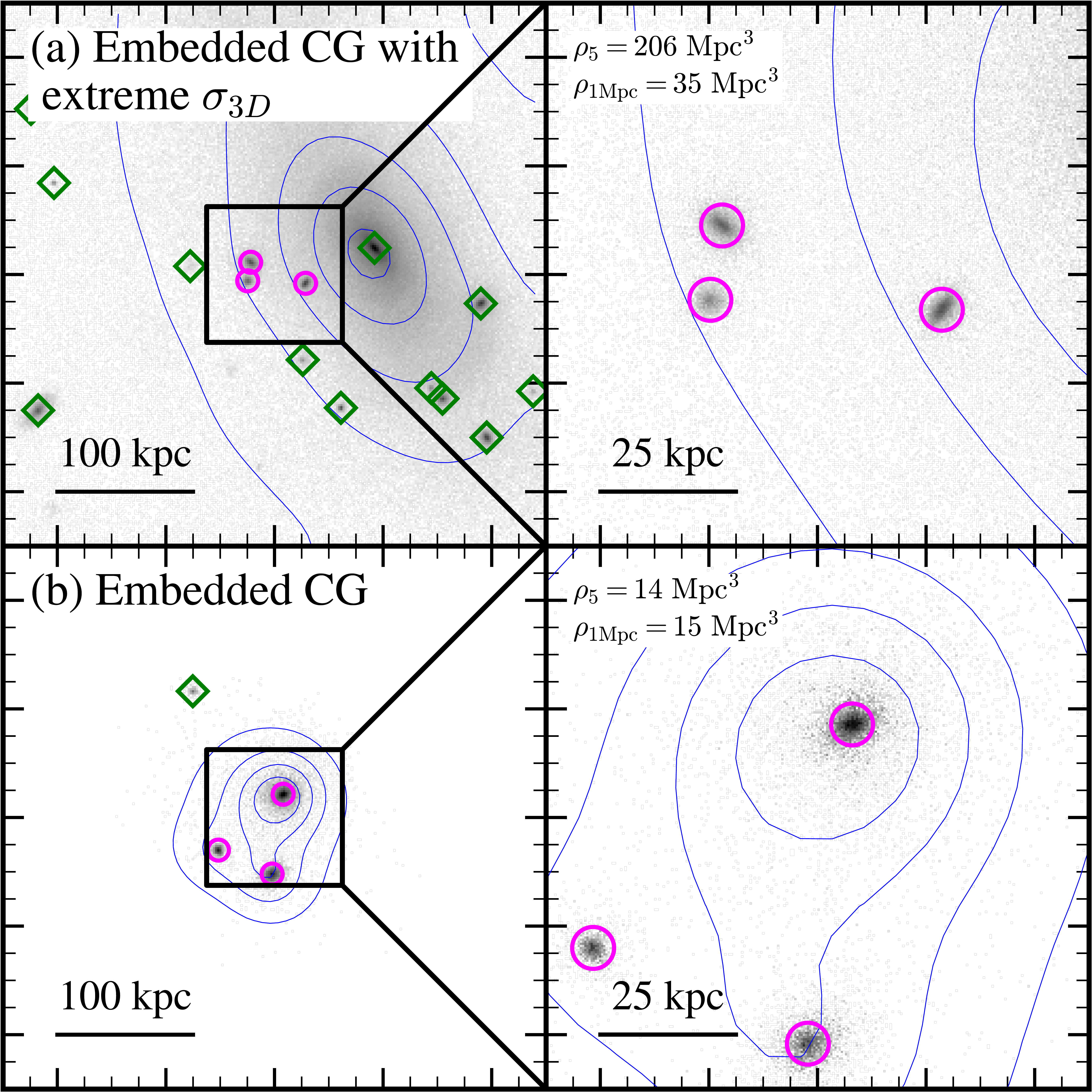}
\caption{(a) The spatial distribution of embedded CGs with extreme $\sigma_{3D}$. The underlying gray density map shows the distribution of stellar particles and blue contours indicate the number density of dark matter particles. Magenta circles mark the center of member subhalos, green diamonds mark the center of neighbor subhalos, orange squares mark the center of non-cosmological origin subhalos. (b) Same as (a), but for example of embedded CGs with that follows the $M_{CG,star}-\sigma_{3D}$ scaling relation.}
\label{fig:CGvisuals}
\end{figure}

Figure \ref{fig:CGvisuals} compares the spatial distributions of galaxies in embedded CGs with extreme $\sigma_{3D}$ (the upper panel), and with normal $\sigma_{3D}$ (the lower panel). The background gray density map and the blue contour map shows the distribution of stellar particles and the number density of dark matter particles, respectively. Magenta circles and green boxes show the position of galaxy subhalos that are members/non-members of the target CGs. The CG members are isolated within a small scale in the three dimensional space, but they are embedded in a nearby massive cluster. In these systems, the strong gravitational potential of the nearby massive clusters inflates the velocity dispersion of the system. 

The systems shown in Figure \ref{fig:CGvisuals} (a) represent systems embedded in extremely high density environments, making them particularly valuable targets for studying the environmental impact on CG evolution. However, despite their extreme environments, these systems are observationally difficult to distinguish from chance alignments when only projected positions and radial velocity differences are available. Their compact appearance in projection plane and line-of-sight velocity differences threshold resemble those of false-positive PPV CGs as we showed in Section \ref{sec:Msig}, highlighting a fundamental limitation of observational CG identification in dense regions.

The full three-dimensional information available in the TNG300 simulation enables us to robustly identify these genuinely compact systems and to characterize their surrounding environments. The CG catalog constructed in simulation provides a unique opportunity to investigate how exceptionally dense environments influence the formation, evolution, and survival of CGs, independent of projection contamination. A detailed analysis of the environmental effects on these abundance and lifetime of CGs will be discussed in our forthcoming paper.

\section{Conclusion} \label{sec:Conclusion}

We construct CG catalogs based on the galaxy subhalos with $M_{*} > 10^{9}~\Mdot$ in the $z=0$ snapshot of the IllustrisTNG-300. We apply the FoF algorithm to identify CGs unlike many previous search for CGs based on the conventional Hickson criteria. In particular, we do not use the Hickson isolation criterion that preferentially removes systems within dense environments. 

We build two CG catalogs. We first identify PPP CGs based on the FoF algorithm with a distance linking length of 73 kpc in physical space. There are 383 PPP CGs in TNG300. These are genuine CGs that provide important testbed for understanding the evolution of CGs in the large-scale structures and the evolution of galaxies in extremely dense environments. 

The second catalog we build is the PPV CG catalog, which is an analogous to the observed CGs. We used the linking lengths: the projected distance linking length and the line-of-sight velocity linking length. Following the identification of observed CGs based on SDSS spectroscopy \citep{Sohn2016}, we use the 73 kpc projected linking length (on the $x\text{-}y$ plane) and the $1000~\kms$ radial velocity (in the $z-$direction) linking length. We account for cosmic expansion by incorporating the line-of-sight expansion velocities of galaxies based on their distances along the $z-$direction. These CGs are useful because they demonstrate how large peculiar velocity difference can confuse the identification of CGs. There are 1666 PPV CGs in TNG300, but majority of them (1328, 80\%) are chance alignments. 

We examine the physical properties of TNG CGs, including the number of members, size, group stellar mass, member galaxy velocity dispersion, and Hubble crossing time. We particularly focus on the group stellar mass ($M_{CG,star}$) to galaxy velocity dispersion relation ($\sigma$). The PPP CGs generally shows a tight relation, $M_{CG,star} \sim \sigma^{3.2}$, with a small fraction of outliers with large velocity dispersion at given group stellar mass. A large fraction of PPV CGs follow the similar relation, but with the much larger scatter in velocity dispersion. More importantly, the fraction of outliers is much larger than for PPP CGs, and most of these outliers are chance alignments. The comparison suggests that identification of outliers based on the $M_{CG,star} - \sigma$ scaling relation effectively selects chance alignments. 

We investigate the completeness of TNG CG identification, particularly focusing on the subhalo identification in TNG300. The TNG subhalo catalog includes non-cosmic origin subhalos that may form during the galaxy mergers that may be frequent in CG environments. Indeed, inclusion of subhalos results in identification of a larger number of CGs. The newly identified CGs have similar physical properties of CGs identified purely based on cosmic origin subhalos. 

We compare the TNG CGs with the observed SDSS CGs. We use the CGs identified by the FoF algorithm from the SDSS volume-limited sample. For a fair comparison, we modify the TNG CG identification by limiting the subhalo mass similar to the magnitude limit applied to the SDSS volume-limited sample. The $M_{CG,star} - \sigma$ relations of the observed CGs are comparable with the PPV CGs; the observed CGs show large scatter in velocity dispersion and the a large fraction of outliers. Since many outliers identified among PPV CGs are chance alignments, the larger fraction of observed CGs indicate that many of them are also chance alignments. 

We finally investigate the density environments of CGs based on two parameters, $\rho_5$ and $\rho_{\mathrm 1Mpc}$. We separate TNG PPP CGs into two groups based on the $\rho_5$ distribution. The embedded CGs show slightly distinctive physical properties; they contain more CGs with larger group stellar mass and the velocity dispersion. These embedded CGs are presumably the part of the larger scale structure (particularly clusters), and the galaxies in the surrounding large scale structure contribute the larger group stellar mass and the larger velocity dispersions. We will discuss the differential evolution of CGs in our forthcoming paper. 

Identification of CGs based on the cosmological simulations provides an important foundation to build the theoretical guidance for evolution of CGs. The systematic search for CGs and tracing them in higher redshift snapshots will offer a precious chance to examine the mysterious evolution of CGs.

\section{Acknowledgments}

We thank the anonymous referee for providing helpful comments that improve the manuscript. This work was supported by Creative-Pioneering Researchers Program through Seoul National University. This work was also supported by the Global-LAMP Program of the National Research Foundation of Korea (NRF) grant funded by the Ministry of Education (No. RS-2023-00301976).

\bibliographystyle{aasjournal}
\bibliography{ms}{}

@ARTICLE{Naiman2018,
       author = {{Naiman}, Jill P. and {Pillepich}, Annalisa and {Springel}, Volker and {Ramirez-Ruiz}, Enrico and {Torrey}, Paul and {Vogelsberger}, Mark and {Pakmor}, R{\"u}diger and {Nelson}, Dylan and {Marinacci}, Federico and {Hernquist}, Lars and {Weinberger}, Rainer and {Genel}, Shy},
        title = "{First results from the IllustrisTNG simulations: a tale of two elements - chemical evolution of magnesium and europium}",
      journal = {\mnras},
         year = 2018,
        month = jun,
       volume = {477},
       number = {1},
        pages = {1206-1224},
          doi = {10.1093/mnras/sty618},
archivePrefix = {arXiv},
       eprint = {1707.03401},
 primaryClass = {astro-ph.GA},
       adsurl = {https://ui.adsabs.harvard.edu/abs/2018MNRAS.477.1206N}
}

@ARTICLE{Pillepich2018,
       author = {{Pillepich}, Annalisa and {Nelson}, Dylan and {Hernquist}, Lars and {Springel}, Volker and {Pakmor}, R{\"u}diger and {Torrey}, Paul and {Weinberger}, Rainer and {Genel}, Shy and {Naiman}, Jill P. and {Marinacci}, Federico and {Vogelsberger}, Mark},
        title = "{First results from the IllustrisTNG simulations: the stellar mass content of groups and clusters of galaxies}",
      journal = {\mnras},
         year = 2018,
        month = mar,
       volume = {475},
       number = {1},
        pages = {648-675},
          doi = {10.1093/mnras/stx3112},
archivePrefix = {arXiv},
       eprint = {1707.03406},
 primaryClass = {astro-ph.GA},
       adsurl = {https://ui.adsabs.harvard.edu/abs/2018MNRAS.475..648P}
}

@ARTICLE{Marinacci2018,
       author = {{Marinacci}, Federico and {Vogelsberger}, Mark and {Pakmor}, R{\"u}diger and {Torrey}, Paul and {Springel}, Volker and {Hernquist}, Lars and {Nelson}, Dylan and {Weinberger}, Rainer and {Pillepich}, Annalisa and {Naiman}, Jill and {Genel}, Shy},
        title = "{First results from the IllustrisTNG simulations: radio haloes and magnetic fields}",
      journal = {\mnras},
         year = 2018,
        month = nov,
       volume = {480},
       number = {4},
        pages = {5113-5139},
          doi = {10.1093/mnras/sty2206},
archivePrefix = {arXiv},
       eprint = {1707.03396},
 primaryClass = {astro-ph.CO},
       adsurl = {https://ui.adsabs.harvard.edu/abs/2018MNRAS.480.5113M}
}

@ARTICLE{Springel2018,
       author = {{Springel}, Volker and {Pakmor}, R{\"u}diger and {Pillepich}, Annalisa and {Weinberger}, Rainer and {Nelson}, Dylan and {Hernquist}, Lars and {Vogelsberger}, Mark and {Genel}, Shy and {Torrey}, Paul and {Marinacci}, Federico and {Naiman}, Jill},
        title = "{First results from the IllustrisTNG simulations: matter and galaxy clustering}",
      journal = {\mnras},
         year = 2018,
        month = mar,
       volume = {475},
       number = {1},
        pages = {676-698},
          doi = {10.1093/mnras/stx3304},
archivePrefix = {arXiv},
       eprint = {1707.03397},
 primaryClass = {astro-ph.GA},
       adsurl = {https://ui.adsabs.harvard.edu/abs/2018MNRAS.475..676S}
}

@ARTICLE{Nelson2018, 
       author = {{Nelson}, Dylan and {Pillepich}, Annalisa and {Springel}, Volker and {Weinberger}, Rainer and {Hernquist}, Lars and {Pakmor}, R{\"u}diger and {Genel}, Shy and {Torrey}, Paul and {Vogelsberger}, Mark and {Kauffmann}, Guinevere and {Marinacci}, Federico and {Naiman}, Jill},
        title = "{First results from the IllustrisTNG simulations: the galaxy colour bimodality}",
      journal = {\mnras},
         year = 2018,
        month = mar,
       volume = {475},
       number = {1},
        pages = {624-647},
          doi = {10.1093/mnras/stx3040},
archivePrefix = {arXiv},
       eprint = {1707.03395},
 primaryClass = {astro-ph.GA},
       adsurl = {https://ui.adsabs.harvard.edu/abs/2018MNRAS.475..624N}
}

@ARTICLE{Springel2001,
       author = {{Springel}, Volker and {White}, Simon D.~M. and {Tormen}, Giuseppe and {Kauffmann}, Guinevere},
        title = "{Populating a cluster of galaxies - I. Results at z=0}",
      journal = {\mnras},
         year = 2001,
        month = dec,
       volume = {328},
       number = {3},
        pages = {726-750},
          doi = {10.1046/j.1365-8711.2001.04912.x},
archivePrefix = {arXiv},
       eprint = {astro-ph/0012055},
 primaryClass = {astro-ph},
       adsurl = {https://ui.adsabs.harvard.edu/abs/2001MNRAS.328..726S}
}

@ARTICLE{Hickson1982,
       author = {{Hickson}, P.},
        title = "{Systematic properties of compact groups of galaxies.}",
      journal = {\apj},
         year = 1982,
        month = apr,
       volume = {255},
        pages = {382-391},
          doi = {10.1086/159838},
       adsurl = {https://ui.adsabs.harvard.edu/abs/1982ApJ...255..382H}
}

@ARTICLE{Hickson1992,
       author = {{Hickson}, Paul and {Mendes de Oliveira}, Claudia and {Huchra}, John P. and {Palumbo}, Giorgio G.},
        title = "{Dynamical Properties of Compact Groups of Galaxies}",
      journal = {\apj},
         year = 1992,
        month = nov,
       volume = {399},
        pages = {353},
          doi = {10.1086/171932},
       adsurl = {https://ui.adsabs.harvard.edu/abs/1992ApJ...399..353H}
}

@ARTICLE{deCarvalho12005,
       author = {{de Carvalho}, R.~R. and {Gon{\c{c}}alves}, T.~S. and {Iovino}, A. and {Kohl-Moreira}, J.~L. and {Gal}, R.~R. and {Djorgovski}, S.~G.},
        title = "{A Catalog of Distant Compact Groups Using the Digitized Second Palomar Observatory Sky Survey}",
      journal = {\aj},
         year = 2005,
        month = aug,
       volume = {130},
       number = {2},
        pages = {425-444},
          doi = {10.1086/430801},
archivePrefix = {arXiv},
       eprint = {astro-ph/0504217},
 primaryClass = {astro-ph},
       adsurl = {https://ui.adsabs.harvard.edu/abs/2005AJ....130..425D}
}

@ARTICLE{Huchra1982,
       author = {{Huchra}, J.~P. and {Geller}, M.~J.},
        title = "{Groups of Galaxies. I. Nearby groups}",
      journal = {\apj},
         year = 1982,
        month = jun,
       volume = {257},
        pages = {423-437},
          doi = {10.1086/160000},
       adsurl = {https://ui.adsabs.harvard.edu/abs/1982ApJ...257..423H}
}

@ARTICLE{Barton1996,
       author = {{Barton}, Elizabeth and {Geller}, Margaret and {Ramella}, Massimo and {Marzke}, Ronald O. and {da Costa}, L. Nicolaci},
        title = "{Compact Group selection From Redshift Surveys}",
      journal = {\aj},
         year = 1996,
        month = sep,
       volume = {112},
        pages = {871},
          doi = {10.1086/118060},
archivePrefix = {arXiv},
       eprint = {astro-ph/9608091},
 primaryClass = {astro-ph},
       adsurl = {https://ui.adsabs.harvard.edu/abs/1996AJ....112..871B}
}

@ARTICLE{Sohn2015,
       author = {{Sohn}, Jubee and {Hwang}, Ho Seong and {Geller}, Margaret J. and {Diaferio}, Antonaldo and {Rines}, Kenneth J. and {Lee}, Myung Gyoon and {Lee}, Gwang-Ho},
        title = "{Compact Groups of Galaxies with Complete Spectroscopic Redshifts in the Local Universe}",
      journal = {Journal of Korean Astronomical Society},
         year = 2015,
        month = dec,
       volume = {48},
       number = {6},
        pages = {381-398},
          doi = {10.5303/JKAS.2015.48.6.381},
archivePrefix = {arXiv},
       eprint = {1601.02646},
 primaryClass = {astro-ph.GA},
       adsurl = {https://ui.adsabs.harvard.edu/abs/2015JKAS...48..381S}
}

@ARTICLE{Sohn2016,
       author = {{Sohn}, Jubee and {Geller}, Margaret J. and {Hwang}, Ho Seong and {Zahid}, H. Jabran and {Lee}, Myung Gyoon},
        title = "{Catalogs of Compact Groups of Galaxies from the Enhanced SDSS DR12}",
      journal = {\apjs},
         year = 2016,
        month = aug,
       volume = {225},
       number = {2},
          eid = {23},
        pages = {23},
          doi = {10.3847/0067-0049/225/2/23},
archivePrefix = {arXiv},
       eprint = {1603.06583},
 primaryClass = {astro-ph.GA},
       adsurl = {https://ui.adsabs.harvard.edu/abs/2016ApJS..225...23S}
}

@ARTICLE{Sohn2017,
       author = {{Sohn}, Jubee and {Zahid}, H. Jabran and {Geller}, Margaret J.},
        title = "{The Velocity Dispersion Function for Quiescent Galaxies in the Local Universe}",
      journal = {\apj},
         year = 2017,
        month = aug,
       volume = {845},
       number = {1},
          eid = {73},
        pages = {73},
          doi = {10.3847/1538-4357/aa7de3},
archivePrefix = {arXiv},
       eprint = {1704.07843},
 primaryClass = {astro-ph.GA},
       adsurl = {https://ui.adsabs.harvard.edu/abs/2017ApJ...845...73S}
}

@ARTICLE{Sohn2021,
       author = {{Sohn}, Jubee and {Geller}, Margaret J. and {Hwang}, Ho Seong and {Fabricant}, Daniel G. and {Moran}, Sean M. and {Utsumi}, Yousuke},
        title = "{The HectoMAP Redshift Survey: First Data Release}",
      journal = {\apj},
         year = 2021,
        month = mar,
       volume = {909},
       number = {2},
          eid = {129},
        pages = {129},
          doi = {10.3847/1538-4357/abd9be},
archivePrefix = {arXiv},
       eprint = {2010.05817},
 primaryClass = {astro-ph.CO},
       adsurl = {https://ui.adsabs.harvard.edu/abs/2021ApJ...909..129S}
}

@ARTICLE{Hartsuiker2020,
       author = {{Hartsuiker}, Len and {Ploeckinger}, Sylvia},
        title = "{Abundance and group coalescence time-scales of compact groups of galaxies in the EAGLE simulation}",
      journal = {\mnras},
         year = 2020,
        month = jan,
       volume = {491},
       number = {1},
        pages = {L66-L71},
          doi = {10.1093/mnrasl/slz171},
archivePrefix = {arXiv},
       eprint = {1911.02025},
 primaryClass = {astro-ph.GA},
       adsurl = {https://ui.adsabs.harvard.edu/abs/2020MNRAS.491L..66H}
}

@ARTICLE{Tempel2016,
       author = {{Tempel}, E. and {Kipper}, R. and {Tamm}, A. and {Gramann}, M. and {Einasto}, M. and {Sepp}, T. and {Tuvikene}, T.},
        title = "{Friends-of-friends galaxy group finder with membership refinement. Application to the local Universe}",
      journal = {\aap},
         year = 2016,
        month = apr,
       volume = {588},
          eid = {A14},
        pages = {A14},
          doi = {10.1051/0004-6361/201527755},
archivePrefix = {arXiv},
       eprint = {1601.01117},
 primaryClass = {astro-ph.CO},
       adsurl = {https://ui.adsabs.harvard.edu/abs/2016A&A...588A..14T}
}

@ARTICLE{Robotham2011,
       author = {{Robotham}, A.~S.~G. and {Norberg}, P. and {Driver}, S.~P. and {Baldry}, I.~K. and {Bamford}, S.~P. and {Hopkins}, A.~M. and {Liske}, J. and {Loveday}, J. and {Merson}, A. and {Peacock}, J.~A. and {Brough}, S. and {Cameron}, E. and {Conselice}, C.~J. and {Croom}, S.~M. and {Frenk}, C.~S. and {Gunawardhana}, M. and {Hill}, D.~T. and {Jones}, D.~H. and {Kelvin}, L.~S. and {Kuijken}, K. and {Nichol}, R.~C. and {Parkinson}, H.~R. and {Pimbblet}, K.~A. and {Phillipps}, S. and {Popescu}, C.~C. and {Prescott}, M. and {Sharp}, R.~G. and {Sutherland}, W.~J. and {Taylor}, E.~N. and {Thomas}, D. and {Tuffs}, R.~J. and {van Kampen}, E. and {Wijesinghe}, D.},
        title = "{Galaxy and Mass Assembly (GAMA): the GAMA galaxy group catalogue (G$^{3}$Cv1)}",
      journal = {\mnras},
         year = 2011,
        month = oct,
       volume = {416},
       number = {4},
        pages = {2640-2668},
          doi = {10.1111/j.1365-2966.2011.19217.x},
archivePrefix = {arXiv},
       eprint = {1106.1994},
 primaryClass = {astro-ph.CO},
       adsurl = {https://ui.adsabs.harvard.edu/abs/2011MNRAS.416.2640R}
}

@ARTICLE{Mamon1986,
       author = {{Mamon}, G.~A.},
        title = "{Are Compact Groups of Galaxies Physically Dense?}",
      journal = {\apj},
         year = 1986,
        month = aug,
       volume = {307},
        pages = {426},
          doi = {10.1086/164431},
       adsurl = {https://ui.adsabs.harvard.edu/abs/1986ApJ...307..426M}
}

@ARTICLE{Beers1990,
       author = {{Beers}, Timothy C. and {Flynn}, Kevin and {Gebhardt}, Karl},
        title = "{Measures of Location and Scale for Velocities in Clusters of Galaxies---A Robust Approach}",
      journal = {\aj},
         year = 1990,
        month = jul,
       volume = {100},
        pages = {32},
          doi = {10.1086/115487},
       adsurl = {https://ui.adsabs.harvard.edu/abs/1990AJ....100...32B}
}

@ARTICLE{Nelson2019,
       author = {{Nelson}, Dylan and {Springel}, Volker and {Pillepich}, Annalisa and {Rodriguez-Gomez}, Vicente and {Torrey}, Paul and {Genel}, Shy and {Vogelsberger}, Mark and {Pakmor}, Ruediger and {Marinacci}, Federico and {Weinberger}, Rainer and {Kelley}, Luke and {Lovell}, Mark and {Diemer}, Benedikt and {Hernquist}, Lars},
        title = "{The IllustrisTNG simulations: public data release}",
      journal = {Computational Astrophysics and Cosmology},
         year = 2019,
        month = may,
       volume = {6},
       number = {1},
          eid = {2},
        pages = {2},
          doi = {10.1186/s40668-019-0028-x},
archivePrefix = {arXiv},
       eprint = {1812.05609},
 primaryClass = {astro-ph.GA},
       adsurl = {https://ui.adsabs.harvard.edu/abs/2019ComAC...6....2N}
}

@ARTICLE{Diaferio1994,
       author = {{Diaferio}, Antonaldo and {Geller}, Margaret J. and {Ramella}, Massimo},
        title = "{The Formation of Compact Groups of Galaxies. I. Optical Properties}",
      journal = {\aj},
         year = 1994,
        month = mar,
       volume = {107},
        pages = {868},
          doi = {10.1086/116900},
       adsurl = {https://ui.adsabs.harvard.edu/abs/1994AJ....107..868D}
}

@ARTICLE{Ramella1994,
       author = {{Ramella}, Massimo and {Diaferio}, Antonaldo and {Geller}, Margaret J. and {Huchra}, John P.},
        title = "{The Birthplace of Compact Groups of Galaxies}",
      journal = {\aj},
         year = 1994,
        month = may,
       volume = {107},
        pages = {1623},
          doi = {10.1086/116971},
       adsurl = {https://ui.adsabs.harvard.edu/abs/1994AJ....107.1623R}
}

@ARTICLE{DiazGimenez2015,
       author = {{D{\'\i}az-Gim{\'e}nez}, Eugenia and {Zandivarez}, Ariel},
        title = "{Where are compact groups in the local Universe?}",
      journal = {\aap},
         year = 2015,
        month = jun,
       volume = {578},
          eid = {A61},
        pages = {A61},
          doi = {10.1051/0004-6361/201425267},
archivePrefix = {arXiv},
       eprint = {1504.02447},
 primaryClass = {astro-ph.GA},
       adsurl = {https://ui.adsabs.harvard.edu/abs/2015A&A...578A..61D}
}

@ARTICLE{Mendel2011,
       author = {{Mendel}, J. Trevor and {Ellison}, Sara L. and {Simard}, Luc and {Patton}, David R. and {McConnachie}, Alan W.},
        title = "{Compact groups in theory and practice - IV. The connection to large-scale structure}",
      journal = {\mnras},
         year = 2011,
        month = dec,
       volume = {418},
       number = {3},
        pages = {1409-1422},
          doi = {10.1111/j.1365-2966.2011.19159.x},
archivePrefix = {arXiv},
       eprint = {1106.1184},
 primaryClass = {astro-ph.CO},
       adsurl = {https://ui.adsabs.harvard.edu/abs/2011MNRAS.418.1409M}
}

@ARTICLE{Barnes1989,
       author = {{Barnes}, Joshua E.},
        title = "{Evolution of compact groups and the formation of elliptical galaxies}",
      journal = {\nat},
         year = 1989,
        month = mar,
       volume = {338},
       number = {6211},
        pages = {123-126},
          doi = {10.1038/338123a0},
       adsurl = {https://ui.adsabs.harvard.edu/abs/1989Natur.338..123B}
}

@ARTICLE{Governato1991,
       author = {{Governato}, Fabio and {Bhatia}, Rajiv and {Chincarini}, Guido},
        title = "{A Long-lasting Compact Group}",
      journal = {\apjl},
         year = 1991,
        month = apr,
       volume = {371},
        pages = {L15},
          doi = {10.1086/185991},
       adsurl = {https://ui.adsabs.harvard.edu/abs/1991ApJ...371L..15G}
}

@ARTICLE{Athanassoula1997,
       author = {{Athanassoula}, E. and {Makino}, J. and {Bosma}, A.},
        title = "{Evolution of compact groups of galaxies - I. Merging rates}",
      journal = {\mnras},
         year = 1997,
        month = apr,
       volume = {286},
       number = {4},
        pages = {825-838},
          doi = {10.1093/mnras/286.4.825},
archivePrefix = {arXiv},
       eprint = {astro-ph/9612223},
 primaryClass = {astro-ph},
       adsurl = {https://ui.adsabs.harvMamard.edu/abs/1997MNRAS.286..825A}
}

@ARTICLE{DiazGimenez2020,
       author = {{D{\'\i}az-Gim{\'e}nez}, E. and {Taverna}, A. and {Zandivarez}, A. and {Mamon}, G.~A.},
        title = "{Compact groups from semi-analytical models of galaxy formation - I. A comparative study of frequency and nature}",
      journal = {\mnras},
         year = 2020,
        month = feb,
       volume = {492},
       number = {2},
        pages = {2588-2605},
          doi = {10.1093/mnras/stz3356},
archivePrefix = {arXiv},
       eprint = {1911.12888},
 primaryClass = {astro-ph.GA},
       adsurl = {https://ui.adsabs.harvard.edu/abs/2020MNRAS.492.2588D}
}

@ARTICLE{Pompei2012,
       author = {{Pompei}, E. and {Iovino}, A.},
        title = "{The DPOSS II distant compact group survey: the EMMI-NTT spectroscopic sample{\ensuremath{\star}}}",
      journal = {\aap},
         year = 2012,
        month = mar,
       volume = {539},
          eid = {A106},
        pages = {A106},
          doi = {10.1051/0004-6361/201118172},
archivePrefix = {arXiv},
       eprint = {1111.1757},
 primaryClass = {astro-ph.CO},
       adsurl = {https://ui.adsabs.harvard.edu/abs/2012A&A...539A.106P}
}

@ARTICLE{Ashman1994,
       author = {{Ashman}, Keith M. and {Bird}, Christina M. and {Zepf}, Stephen E.},
        title = "{Detecting Bimodality in Astronomical Datasets}",
      journal = {\aj},
         year = 1994,
        month = dec,
       volume = {108},
        pages = {2348},
          doi = {10.1086/117248},
archivePrefix = {arXiv},
       eprint = {astro-ph/9408030},
 primaryClass = {astro-ph},
       adsurl = {https://ui.adsabs.harvard.edu/abs/1994AJ....108.2348A}
}

@ARTICLE{Taverna2024,
       author = {{Taverna}, A. and {D{\'\i}az-Gim{\'e}nez}, E. and {Zandivarez}, A. and {Mart{\'\i}nez}, H.~J. and {Ruiz}, A.~N.},
        title = "{Compact groups from semi-analytical models of galaxy formation - V. Their assembly channels as a function of the environment}",
      journal = {\mnras},
         year = 2024,
        month = jan,
       volume = {527},
       number = {3},
        pages = {4821-4833},
          doi = {10.1093/mnras/stad3512},
archivePrefix = {arXiv},
       eprint = {2311.06721},
 primaryClass = {astro-ph.GA},
       adsurl = {https://ui.adsabs.harvard.edu/abs/2024MNRAS.527.4821T}
}

@ARTICLE{MendesdeOliveira1991,
       author = {{Mendes de Oliveira}, Claudia and {Hickson}, Paul},
        title = "{The Luminosity Function of Compact Groups of Galaxies}",
      journal = {\apj},
         year = 1991,
        month = oct,
       volume = {380},
        pages = {30},
          doi = {10.1086/170559},
       adsurl = {https://ui.adsabs.harvard.edu/abs/1991ApJ...380...30M}
}

@ARTICLE{Duplancic2013,
       author = {{Duplancic}, Fernanda and {O'Mill}, Ana Laura and {Lambas}, Diego G. and {Sodr{\'e}}, Laerte and {Alonso}, Sol},
        title = "{Galaxy triplets in Sloan Digital Sky Survey Data Release 7 - II. A connection with compact groups?}",
      journal = {\mnras},
         year = 2013,
        month = aug,
       volume = {433},
       number = {4},
        pages = {3547-3558},
          doi = {10.1093/mnras/stt985},
archivePrefix = {arXiv},
       eprint = {1306.0891},
 primaryClass = {astro-ph.CO},
       adsurl = {https://ui.adsabs.harvard.edu/abs/2013MNRAS.433.3547D}
}

@ARTICLE{Stetson1987,
       author = {{Stetson}, Peter B.},
        title = "{DAOPHOT: A Computer Program for Crowded-Field Stellar Photometry}",
      journal = {\pasp},
         year = 1987,
        month = mar,
       volume = {99},
        pages = {191},
          doi = {10.1086/131977},
       adsurl = {https://ui.adsabs.harvard.edu/abs/1987PASP...99..191S}
}

@ARTICLE{Zahid2016,
       author = {{Zahid}, H. Jabran and {Geller}, Margaret J. and {Fabricant}, Daniel G. and {Hwang}, Ho Seong},
        title = "{The Scaling of Stellar Mass and Central Stellar Velocity Dispersion for Quiescent Galaxies at z<0.7}",
      journal = {\apj},
         year = 2016,
        month = dec,
       volume = {832},
       number = {2},
          eid = {203},
        pages = {203},
          doi = {10.3847/0004-637X/832/2/203},
archivePrefix = {arXiv},
       eprint = {1607.04275},
 primaryClass = {astro-ph.GA},
       adsurl = {https://ui.adsabs.harvard.edu/abs/2016ApJ...832..203Z}
}

@ARTICLE{Evrard2008,
       author = {{Evrard}, A.~E. and {Bialek}, J. and {Busha}, M. and {White}, M. and {Habib}, S. and {Heitmann}, K. and {Warren}, M. and {Rasia}, E. and {Tormen}, G. and {Moscardini}, L. and {Power}, C. and {Jenkins}, A.~R. and {Gao}, L. and {Frenk}, C.~S. and {Springel}, V. and {White}, S.~D.~M. and {Diemand}, J.},
        title = "{Virial Scaling of Massive Dark Matter Halos: Why Clusters Prefer a High Normalization Cosmology}",
      journal = {\apj},
         year = 2008,
        month = jan,
       volume = {672},
       number = {1},
        pages = {122-137},
          doi = {10.1086/521616},
archivePrefix = {arXiv},
       eprint = {astro-ph/0702241},
 primaryClass = {astro-ph},
       adsurl = {https://ui.adsabs.harvard.edu/abs/2008ApJ...672..122E}
}

@ARTICLE{Rines2016,
       author = {{Rines}, Kenneth J. and {Geller}, Margaret J. and {Diaferio}, Antonaldo and {Hwang}, Ho Seong},
        title = "{HeCS-SZ: The Hectospec Survey of Sunyaev-Zeldovich-selected Clusters}",
      journal = {\apj},
         year = 2016,
        month = mar,
       volume = {819},
       number = {1},
          eid = {63},
        pages = {63},
          doi = {10.3847/0004-637X/819/1/63},
archivePrefix = {arXiv},
       eprint = {1507.08289},
 primaryClass = {astro-ph.CO},
       adsurl = {https://ui.adsabs.harvard.edu/abs/2016ApJ...819...63R}
}

@ARTICLE{Arnouts1999,
       author = {{Arnouts}, S. and {Cristiani}, S. and {Moscardini}, L. and {Matarrese}, S. and {Lucchin}, F. and {Fontana}, A. and {Giallongo}, E.},
        title = "{Measuring and modelling the redshift evolution of clustering: the Hubble Deep Field North}",
      journal = {\mnras},
         year = 1999,
        month = dec,
       volume = {310},
       number = {2},
        pages = {540-556},
          doi = {10.1046/j.1365-8711.1999.02978.x},
archivePrefix = {arXiv},
       eprint = {astro-ph/9902290},
 primaryClass = {astro-ph},
       adsurl = {https://ui.adsabs.harvard.edu/abs/1999MNRAS.310..540A}
}

@ARTICLE{Ilbert2006,
       author = {{Ilbert}, O. and {Arnouts}, S. and {McCracken}, H.~J. and {Bolzonella}, M. and {Bertin}, E. and {Le F{\`e}vre}, O. and {Mellier}, Y. and {Zamorani}, G. and {Pell{\`o}}, R. and {Iovino}, A. and {Tresse}, L. and {Le Brun}, V. and {Bottini}, D. and {Garilli}, B. and {Maccagni}, D. and {Picat}, J.~P. and {Scaramella}, R. and {Scodeggio}, M. and {Vettolani}, G. and {Zanichelli}, A. and {Adami}, C. and {Bardelli}, S. and {Cappi}, A. and {Charlot}, S. and {Ciliegi}, P. and {Contini}, T. and {Cucciati}, O. and {Foucaud}, S. and {Franzetti}, P. and {Gavignaud}, I. and {Guzzo}, L. and {Marano}, B. and {Marinoni}, C. and {Mazure}, A. and {Meneux}, B. and {Merighi}, R. and {Paltani}, S. and {Pollo}, A. and {Pozzetti}, L. and {Radovich}, M. and {Zucca}, E. and {Bondi}, M. and {Bongiorno}, A. and {Busarello}, G. and {de La Torre}, S. and {Gregorini}, L. and {Lamareille}, F. and {Mathez}, G. and {Merluzzi}, P. and {Ripepi}, V. and {Rizzo}, D. and {Vergani}, D.},
        title = "{Accurate photometric redshifts for the CFHT legacy survey calibrated using the VIMOS VLT deep survey}",
      journal = {\aap},
         year = 2006,
        month = oct,
       volume = {457},
       number = {3},
        pages = {841-856},
          doi = {10.1051/0004-6361:20065138},
archivePrefix = {arXiv},
       eprint = {astro-ph/0603217},
 primaryClass = {astro-ph},
       adsurl = {https://ui.adsabs.harvard.edu/abs/2006A&A...457..841I}
}

@ARTICLE{FloresFreitas2024,
       author = {{Flores-Freitas}, Rodrigo and {Trevisan}, Marina and {M{\"u}ckler}, Mait{\^e} and {Mamon}, Gary A. and {Schnorr-M{\"u}ller}, Allan and {Bootz}, Vitor},
        title = "{Compact groups of dwarf galaxies in TNG50: late hierarchical assembly and delayed stellar build-up in the low-mass regime}",
      journal = {\mnras},
         year = 2024,
        month = mar,
       volume = {528},
       number = {4},
        pages = {5804-5824},
          doi = {10.1093/mnras/stae367},
archivePrefix = {arXiv},
       eprint = {2401.13252},
 primaryClass = {astro-ph.GA},
       adsurl = {https://ui.adsabs.harvard.edu/abs/2024MNRAS.528.5804F}
}

@ARTICLE{Walker2013,
       author = {{Walker}, Lisa May and {Butterfield}, Natalie and {Johnson}, Kelsey and {Zucker}, Catherine and {Gallagher}, Sarah and {Konstantopoulos}, Iraklis and {Zabludoff}, Ann and {Hornschemeier}, Ann E. and {Tzanavaris}, Panayiotis and {Charlton}, Jane C.},
        title = "{The Optical Green Valley versus Mid-infrared Canyon in Compact Groups}",
      journal = {\apj},
         year = 2013,
        month = oct,
       volume = {775},
       number = {2},
          eid = {129},
        pages = {129},
          doi = {10.1088/0004-637X/775/2/129},
archivePrefix = {arXiv},
       eprint = {1307.6559},
 primaryClass = {astro-ph.CO},
       adsurl = {https://ui.adsabs.harvard.edu/abs/2013ApJ...775..129W}
}

@ARTICLE{McConnachie2009,
       author = {{McConnachie}, Alan W. and {Patton}, David R. and {Ellison}, Sara L. and {Simard}, Luc},
        title = "{Compact groups in theory and practice - III. Compact groups of galaxies in the Sixth Data Release of the Sloan Digital Sky Survey}",
      journal = {\mnras},
         year = 2009,
        month = may,
       volume = {395},
       number = {1},
        pages = {255-268},
          doi = {10.1111/j.1365-2966.2008.14340.x},
archivePrefix = {arXiv},
       eprint = {0812.1580},
 primaryClass = {astro-ph},
       adsurl = {https://ui.adsabs.harvard.edu/abs/2009MNRAS.395..255M}
}

@ARTICLE{Lee2017,
       author = {{Lee}, Gwang-Ho and {Hwang}, Ho Seong and {Sohn}, Jubee and {Lee}, Myung Gyoon},
        title = "{The Fastest Galaxy Evolution in an Unbiased Compact Group Sample with WISE}",
      journal = {\apj},
         year = 2017,
        month = feb,
       volume = {835},
       number = {2},
          eid = {280},
        pages = {280},
          doi = {10.3847/1538-4357/835/2/280},
archivePrefix = {arXiv},
       eprint = {1701.01102},
 primaryClass = {astro-ph.GA},
       adsurl = {https://ui.adsabs.harvard.edu/abs/2017ApJ...835..280L}
}

@ARTICLE{VerdesMontenegro1998,
       author = {{Verdes-Montenegro}, L. and {Yun}, M.~S. and {Perea}, J. and {del Olmo}, A. and {Ho}, P.~T.~P.},
        title = "{Effects of Interaction-induced Activities in Hickson Compact Groups: CO and Far-Infrared Study}",
      journal = {\apj},
         year = 1998,
        month = apr,
       volume = {497},
       number = {1},
        pages = {89-107},
          doi = {10.1086/305454},
archivePrefix = {arXiv},
       eprint = {astro-ph/9711127},
 primaryClass = {astro-ph},
       adsurl = {https://ui.adsabs.harvard.edu/abs/1998ApJ...497...89V}
}

@ARTICLE{VerdesMontenegro2001,
       author = {{Verdes-Montenegro}, L. and {Yun}, M.~S. and {Williams}, B.~A. and {Huchtmeier}, W.~K. and {Del Olmo}, A. and {Perea}, J.},
        title = "{Where is the neutral atomic gas in Hickson groups?}",
      journal = {\aap},
         year = 2001,
        month = oct,
       volume = {377},
        pages = {812-826},
          doi = {10.1051/0004-6361:20011127},
archivePrefix = {arXiv},
       eprint = {astro-ph/0108223},
 primaryClass = {astro-ph},
       adsurl = {https://ui.adsabs.harvard.edu/abs/2001A&A...377..812V}
}

@ARTICLE{Rasmussen2008,
       author = {{Rasmussen}, Jesper and {Ponman}, Trevor J. and {Verdes-Montenegro}, Lourdes and {Yun}, Min S. and {Borthakur}, Sanchayeeta},
        title = "{Galaxy evolution in Hickson compact groups: the role of ram-pressure stripping and strangulation}",
      journal = {\mnras},
         year = 2008,
        month = aug,
       volume = {388},
       number = {3},
        pages = {1245-1264},
          doi = {10.1111/j.1365-2966.2008.13451.x},
archivePrefix = {arXiv},
       eprint = {0805.1709},
 primaryClass = {astro-ph},
       adsurl = {https://ui.adsabs.harvard.edu/abs/2008MNRAS.388.1245R}
}

@ARTICLE{Bitsakis2016,
       author = {{Bitsakis}, T. and {Dultzin}, D. and {Ciesla}, L. and {D{\'\i}az-Santos}, T. and {Appleton}, P.~N. and {Charmandaris}, V. and {Krongold}, Y. and {Guillard}, P. and {Alatalo}, K. and {Zezas}, A. and {Gonz{\'a}lez}, J. and {Lanz}, L.},
        title = "{Studying the evolution of galaxies in compact groups over the past 3 Gyr - II. The importance of environment in the suppression of star formation}",
      journal = {\mnras},
         year = 2016,
        month = jun,
       volume = {459},
       number = {1},
        pages = {957-970},
          doi = {10.1093/mnras/stw686},
archivePrefix = {arXiv},
       eprint = {1603.06007},
 primaryClass = {astro-ph.GA},
       adsurl = {https://ui.adsabs.harvard.edu/abs/2016MNRAS.459..957B}
}

@ARTICLE{DiazGimenez2018,
       author = {{D{\'\i}az-Gim{\'e}nez}, Eugenia and {Zandivarez}, Ariel and {Taverna}, Antonela},
        title = "{Improving Hickson-like compact group finders in redshift surveys: an implementation in the SDSS}",
      journal = {\aap},
         year = 2018,
        month = oct,
       volume = {618},
          eid = {A157},
        pages = {A157},
          doi = {10.1051/0004-6361/201833329},
archivePrefix = {arXiv},
       eprint = {1808.10051},
 primaryClass = {astro-ph.GA},
       adsurl = {https://ui.adsabs.harvard.edu/abs/2018A&A...618A.157D}
}

@ARTICLE{Lee2004,
       author = {{Lee}, Brian C. and {Allam}, Sahar S. and {Tucker}, Douglas L. and {Annis}, James and {Johnston}, David E. and {Scranton}, Ryan and {Acebo}, Yamina and {Bahcall}, Neta A. and {Bartelmann}, Matthias and {B{\"o}hringer}, Hans and {Ellman}, Nancy and {Grebel}, Eva K. and {Infante}, Leopoldo and {Loveday}, Jon and {McKay}, Timothy A. and {Prada}, Francisco and {Schneider}, Donald P. and {Stoughton}, Chris and {Szalay}, Alexander S. and {Vogeley}, Michael S. and {Voges}, Wolfgang and {Yanny}, Brian},
        title = "{A Catalog of Compact Groups of Galaxies in the SDSS Commissioning Data}",
      journal = {\aj},
         year = 2004,
        month = apr,
       volume = {127},
       number = {4},
        pages = {1811-1859},
          doi = {10.1086/382236},
archivePrefix = {arXiv},
       eprint = {astro-ph/0312553},
 primaryClass = {astro-ph},
       adsurl = {https://ui.adsabs.harvard.edu/abs/2004AJ....127.1811L}
}

@ARTICLE{Zheng2020,
       author = {{Zheng}, Yun-Liang and {Shen}, Shi-Yin},
        title = "{Compact Groups of Galaxies in Sloan Digital Sky Survey and LAMOST Spectral Survey. I. The Catalogs}",
      journal = {\apjs},
         year = 2020,
        month = jan,
       volume = {246},
       number = {1},
          eid = {12},
        pages = {12},
          doi = {10.3847/1538-4365/ab5c26},
archivePrefix = {arXiv},
       eprint = {1911.11478},
 primaryClass = {astro-ph.GA},
       adsurl = {https://ui.adsabs.harvard.edu/abs/2020ApJS..246...12Z}
}

@ARTICLE{Prandoni1994,
       author = {{Prandoni}, I. and {Iovino}, A. and {MacGillivray}, H.~T.},
        title = "{Automated Search For Compact Groups of Galaxies in The Southern Sky}",
      journal = {\aj},
         year = 1994,
        month = apr,
       volume = {107},
        pages = {1235},
          doi = {10.1086/116936},
       adsurl = {https://ui.adsabs.harvard.edu/abs/1994AJ....107.1235P}
}

@ARTICLE{Zandivarez2024,
       author = {{Zandivarez}, A. and {D{\'\i}az-Gim{\'e}nez}, E. and {Taverna}, A. and {Rodriguez}, F. and {Merch{\'a}n}, M.},
        title = "{Compact groups of galaxies in GAMA: Probing the densest minor systems at intermediate redshifts}",
      journal = {\aap},
         year = 2024,
        month = nov,
       volume = {691},
          eid = {A6},
        pages = {A6},
          doi = {10.1051/0004-6361/202451471},
archivePrefix = {arXiv},
       eprint = {2408.07031},
 primaryClass = {astro-ph.GA},
       adsurl = {https://ui.adsabs.harvard.edu/abs/2024A&A...691A...6Z}
}

@ARTICLE{Carnevali1981,
       author = {{Carnevali}, P. and {Cavaliere}, A. and {Santangelo}, P.},
        title = "{Merging instability in groups of galaxies}",
      journal = {\apj},
         year = 1981,
        month = oct,
       volume = {249},
        pages = {449-461},
          doi = {10.1086/159305},
       adsurl = {https://ui.adsabs.harvard.edu/abs/1981ApJ...249..449C}
}

@ARTICLE{Celiz2025,
       author = {{Celiz}, Bruno M. and {Benavides}, Jos{\'e} A. and {Abadi}, Mario G.},
        title = "{Compact groups of galaxies in the TNG100 simulation}",
      journal = {\aap},
         year = 2025,
        month = sep,
       volume = {702},
          eid = {A22},
        pages = {A22},
          doi = {10.1051/0004-6361/202555375},
archivePrefix = {arXiv},
       eprint = {2508.13293},
 primaryClass = {astro-ph.GA},
       adsurl = {https://ui.adsabs.harvard.edu/abs/2025A&A...702A..22C}
}

@ARTICLE{Ebeling1994,
       author = {{Ebeling}, Harald and {Voges}, Wolfgang and {Boehringer}, Hans},
        title = "{X-Ray Emission from Hickson's Compact Groups of Galaxies: Results from the ROSAT All-Sky Survey}",
      journal = {\apj},
         year = 1994,
        month = nov,
       volume = {436},
        pages = {44},
          doi = {10.1086/174879},
       adsurl = {https://ui.adsabs.harvard.edu/abs/1994ApJ...436...44E}
}

@ARTICLE{Iovino2002,
       author = {{Iovino}, Angela},
        title = "{Detecting Fainter Compact Groups: Results from a New Automated Algorithm}",
      journal = {\aj},
         year = 2002,
        month = nov,
       volume = {124},
       number = {5},
        pages = {2471-2489},
          doi = {10.1086/343059},
       adsurl = {https://ui.adsabs.harvard.edu/abs/2002AJ....124.2471I}
}

@ARTICLE{Iovino2003,
       author = {{Iovino}, A. and {de Carvalho}, R.~R. and {Gal}, R.~R. and {Odewahn}, S.~C. and {Lopes}, P.~A.~A. and {Mahabal}, A. and {Djorgovski}, S.~G.},
        title = "{A New Sample of Distant Compact Groups from the Digitized Second Palomar Observatory Sky Survey}",
      journal = {\aj},
         year = 2003,
        month = apr,
       volume = {125},
       number = {4},
        pages = {1660-1681},
          doi = {10.1086/373999},
       adsurl = {https://ui.adsabs.harvard.edu/abs/2003AJ....125.1660I}
}

@ARTICLE{DiazGimenez2012,
       author = {{D{\'\i}az-Gim{\'e}nez}, Eugenia and {Mamon}, Gary A. and {Pacheco}, Marcela and {Mendes de Oliveira}, Claudia and {Alonso}, M. Victoria},
        title = "{Compact groups of galaxies selected by stellar mass: the 2MASS compact group catalogue}",
      journal = {\mnras},
         year = 2012,
        month = oct,
       volume = {426},
       number = {1},
        pages = {296-316},
          doi = {10.1111/j.1365-2966.2012.21705.x},
archivePrefix = {arXiv},
       eprint = {1207.2196},
 primaryClass = {astro-ph.CO},
       adsurl = {https://ui.adsabs.harvard.edu/abs/2012MNRAS.426..296D}
}

@ARTICLE{Ponman1996,
       author = {{Ponman}, T.~J. and {Bourner}, P.~D.~J. and {Ebeling}, H. and {B{\"o}hringer}, H.},
        title = "{A ROSAT survey of Hickson's compact galaxy groups.}",
      journal = {\mnras},
         year = 1996,
        month = dec,
       volume = {283},
        pages = {690-708},
          doi = {10.1093/mnras/283.2.690},
archivePrefix = {arXiv},
       eprint = {astro-ph/9607114},
 primaryClass = {astro-ph},
       adsurl = {https://ui.adsabs.harvard.edu/abs/1996MNRAS.283..690P}
}

@ARTICLE{Zandivarez2022,
       author = {{Zandivarez}, A. and {D{\'\i}az-Gim{\'e}nez}, E. and {Taverna}, A.},
        title = "{The influence of Hickson-like compact group environment on galaxy luminosities}",
      journal = {\mnras},
         year = 2022,
        month = jul,
       volume = {514},
       number = {1},
        pages = {1231-1248},
          doi = {10.1093/mnras/stac1374},
archivePrefix = {arXiv},
       eprint = {2205.07341},
 primaryClass = {astro-ph.GA},
       adsurl = {https://ui.adsabs.harvard.edu/abs/2022MNRAS.514.1231Z}
}

\end{document}